\documentclass[twocolumn]{aastex631}%
\usepackage{multirow}
\usepackage{diagbox}

\usepackage{soul}

\begin{document}


\title{The Galaxy Stellar Mass–SFR–Size Relation in EAGLE, TNG100, and Observations}

\author{Jiani Chu}
\affiliation{Department of Astronomy, Tsinghua University, Beijing, 100084, China. }

\author{Dandan Xu}
\affiliation{Department of Astronomy, Tsinghua University, Beijing, 100084, China. }

\author{Enci Wang}
\affiliation{Department of Astronomy, University of Science and Technology of China, Hefei 230026, China}
\affiliation{School of Astronomy and Space Science, University of Science and Technology of China, Hefei 230026, China}

\author{Stijn Wuyts}
\affiliation{Department of Physics, University of Bath, Claverton Down, Bath BA2 7AY, UK}

\correspondingauthor{Jiani Chu \& Dandan Xu}
\email{zjn20@mails.tsinghua.edu.cn}
\email{dandanxu@tsinghua.edu.cn}







\begin{abstract}

Stellar mass, size, and star formation rate (SFR) are fundamental properties that encode the structural and evolutionary states of galaxies. Observations reveal a mass--SFR--size relation whereby galaxies become more compact both above and below the ridge of the star-forming main sequence (SFMS), linking galaxy structure to star formation activity. We investigate this relation by comparing galaxies from two cosmological hydrodynamical simulations, \textsc{EAGLE} and \textsc{TNG100}, with observational samples from \textsc{SDSS} and \textsc{CANDELS} over three redshift intervals ($0 \leq z < 0.2$, $0.5 \leq z < 1.5$, and $1.5 \leq z < 2.5$). Both simulations reproduce the observed trend that galaxy sizes decrease with increasing offset away from the SFMS. This trend, however, weakens and is not detected in the observational sample at $1.5 < z < 2.5$, likely due to increased measurement uncertainties. In contrast, the trend persists in both simulations up to $z = 2.5$. Across all redshifts, \textsc{EAGLE} predicts a stronger size dependence on SFMS offset than observed, whereas \textsc{TNG100} exhibits a weaker dependence. We discuss how this mass–SFR–size relation can be understood in terms of different time variability in star formation rate across the SFMS.

\end{abstract}



\section{Introduction}

Stellar mass, size, and star formation rate (SFR) are among the most fundamental properties governing the formation and evolution of galaxies. While the stellar mass–size and stellar mass–SFR relations have been extensively studied, the combined interplay among these three quantities remains less well understood.

Star-forming galaxies are observed to populate a well-defined ``star-forming main sequence'' (SFMS) in the SFR–stellar mass plane across a wide range of redshifts, characterized by a tight correlation and an intrinsic scatter of $\sim$0.3 dex \citep{Brinchmann2004, Noeske2007, Daddi2007, Elbaz2007, Whitaker2012, Salim2017, Pannella2015, Pearson2018, Leslie2020, Popesso2023}. This tight relation of star-forming galaxies is generally understood to arise from the interplay among gas inflow, outflow, and star formation \citep{Tacchella2013, Dekel2013, Lilly2013, Forbes2014, Schreiber2015, Wang2019,  Wang2021}. The normalization of the SFMS exhibits strong redshift evolution, with a systematic decline toward the present epoch \citep{Speagle2014, Song2025}. In the local universe, a substantial fraction of galaxies have ceased forming stars and populate the so-called ``red sequence'' below the SFMS \citep{Kauffmann2003, Faber2007, Franx_2008, Schaye2010, Peng_2010, Peng2015Natur, vanDokkum2015, Bluck2016, Whitaker2017, Mosleh2017}. While SFR generally increases with stellar mass among star-forming systems, the most massive galaxies are predominantly quiescent and dominated by old stellar populations.

The galaxy stellar mass–size relation has also been extensively studied across a wide range of redshifts \citep[e.g.][]{Daddi2005, Trujillo2007, Cimatti2008, McGrath_2008, vanDokkum2008, Szomoru2011, Newman2012, Barro2013, Dullo_2013, Poggianti2013, Shankar2013, Carollo2013, Cassata2013}. Observationally, galaxy sizes are found to increase with cosmic time, with more massive galaxies generally exhibiting larger sizes than their lower-mass counterparts. At fixed stellar mass, star-forming galaxies are systematically larger than quiescent systems \citep[e.g.][]{Shen2003, Ferguson2004, Trujillo2006, Elmegreen2007, Franx_2008, Toft2009, Bezanson2009, Williams2010, Szomoru2011, Wuyts2011, Dutton2011, Baldry2012, Mosleh2012, Ono2013, vanDerWel2014, Lange2015}. In addition, measured galaxy sizes show a systematic dependence on observed wavelength, reflecting variations in stellar populations and dust attenuation \citep[e.g.][]{Kelvin2012, Bond2014, Jia2024, vanderWel2024, Martorano2026}.

In a classical scenario,  galaxy size is thought to be closely related to the host dark matter halo spin, which is a result of tidal torque theory in the linear regime (e.g., \citealt{Peebles_1969, doroshkevich1970, Fall_Efstathiou_1980, white1984, barnes1987, Catelan_Theuns_1996}). During halo collapse and galaxy formation, gas radiates and cools while conserving specific angular momentum, resulting in flat rotating gaseous disks from which stars form. Angular momentum is thus transferred from gas to stars, building up spin and size of a galaxy (e.g. \citealt{Fall1980, Efstathiou1982, White1978, White1991, Mo1998, Dekel2014, Schaefer_2009_Review, Wang2022}). 
In reality, galaxies as dynamical systems suffer from various kinds of gravitational instability which influences the angular momentum and size of a galaxy. In particular, this instability is tightly coupled with local specific angular momentum across various morphological types \citep{Romeo2018,Romeo2023}. 
Not only so, their merging history and interacting environment also add complications to the above-mentioned classical halo collapse and disk formation scenario, in particular, galaxies are expected to grow significantly via galaxy mergers (e.g., \citealt{Naab2009, Hopkins2010, Oser2010, Oogi2013, Furong2017}). 
As a consequence, dark matter, gas and stars will all alter their angular momentum distributions and experience angular momentum exchange (e.g., \citealt{Zjupa2017,DeFelippis2017}), on its turn directly impacting galaxy size.


It is worth noting that galaxy size and SFR are not independent. On the one hand, the angular momentum environment of the circumgalactic gas directly affects the cold gas accretion in star-forming galaxies (\citealt{WangSen2025}), which both builds up stellar disks and enhances star formation activity \citep{Gui2025}. On the other hand, feedback processes regulate SFR with spatially varying efficiency, which can in turn modify the apparent sizes of stellar disks (e.g., \citealt{Fan2008, Fan2010, Dubois2013, vanDokkum2014, Governato2004, Sales2010, Brook2012, Brook2012b, McCarthy2012, Aumer2013, Munshi2013, Hopkins2014, Marinacci2014, Crain2015}). 
Therefore galaxy size serves as an important observational feature to constrain various possible feedback mechanisms \citep{Ma2024, Jia2025}. In addition, the size of a galaxy (and in detail both the extent of its bulge and disk) also affects star forming activities by stabilizing gas against collapsing to form stars (see \citealt{Toomre1964, Martig2009}). As a result, more compact galaxies generally have experienced more intense starbursts in their early history and become more quiescent at later times. In contrast, more extended galaxies have always had relatively mild star forming activities and thus manage to maintain active star formation until today.

In this regard, \citet{Wuyts2011} reported a subtle and detailed connection between stellar mass, size (or morphology), and SFR, using Hubble Space Telescope imaging and a Herschel-calibrated ladder of SFR indicators out to 
$z=2.5$ (see also \citealt{he2025symmetryfundamentalparametersgalaxies}).
The study clearly demonstrated that as galaxies move vertically across the SFMS in the SFR-mass plane, the further they deviate away from the SFMS ridge, the smaller their sizes are and the more compact they become. Specifically, galaxies on the SFMS ridge are systematically bigger and more extended (with Sersic index $n\sim1$) while galaxies living further away from the ridge are smaller and more compact (with $n\geqslant2$, more precisely, $n\sim 2$ for above SFMS, $n\sim4$ for below SFMS, see Figure 2 therein). Such a trend directly shows that galaxy structure/morphology is not solely a function of galaxy mass and SFR, but the offset in SFR to the SFMS ridge. This further reveals a possibly deeper connection between stellar population and galaxy structure: in particular, a plausible connection between the transition of galaxy SFR (thus resulting in star-forming and quenched galaxy populations) and the transition of galaxy morphology (thus resulting in more extended disk galaxies and more compact bulgy galaxies), which already exists at least to $z\sim 1.5$.

In this study, we aim at investigating the existence of such a mass-SFR-size relation in the latest cosmological simulations. To do so, we utilize as a reference the data from both SDSS and CANDELS \citep{Grogin2011, Koekemoer2011} 
galaxies to reproduce the observed SFR-size-mass relation up to redshift $z=2.5$. In particular, we compare results from two simulation suites to these observations. The first one is EAGLE simulation \citep{Schaye2015, McAlpine2016}, and the second one is TNG100\footnote{\url{www.tng-project.org}} (\citealt{Genel_2018MNRAS_TNG_SizeEvo, Nelson18_TNGcolor, Pillepich_et_al.(2018b), Springel_et_al.(2018), Marinacci18_TNGmagnetic, Naiman_et_al.(2018)}) simulation. Both simulations have been shown to roughly produce the observed SFMS (see e.g., \citealt{Furong2015, Pillepich_et_al.(2018b), Donnari2018}) and the mass-size relation (see e.g., \citealt{Furong2017, deGraaff2022_EAGLEmasssize, Genel_2018MNRAS_TNG_SizeEvo}). 
We note that agreements between simulations and observations in the stellar mass–SFR and stellar mass–size relations individually do not necessarily imply consistency in the joint relation among all three quantities, given their intrinsic interconnections. In this study, we therefore extend the comparison into the joint mass-SFR-size relation. 


The paper is organized as follows: In Section\,\ref{sec:obs_data}, we introduce the observational data that we adopt as a reference. In section\,\ref{sec:sim_data}, we introduce two simulation samples from both EAGLE and TNG100 simulations. In order to remove the mass dependence, in this study, we focus on the relative sizes of galaxies at any given stellar mass, in section\,\ref{sec:DRe} we present the methodology to calculate the relative sizes of main-sequence galaxies with respect to systems of the same stellar mass and redshift for a given dataset. In section\,\ref{sec:results}, we present a detailed comparison in the stellar-SFR-size relation between observations and the two simulations. Finally, in section\,\ref{sec:conclusion}, we present the overall conclusion and a discussion on how the observed relation can be explained in terms of time variability in star formation. Throughout this study, we adopt a flat $\Lambda$CDM cosmology based on the Planck results (\citealt{Planck_Collaboration2016}), i.e., with a total matter density $\Omega_{\rm m} = 0.3089$, a baryonic matter density $\Omega_{\rm b} = 0.0486$, and a Hubble constant $h = H_0/(100\,{\rm km s}^{-1} {\rm Mpc^{-1}}) = 0.6774$.

\section{Observational data} \label{sec:obs_data}

In the analysis, we follow the same redshift binning scheme as adopted by \citet{Wuyts2011}, dividing galaxies into three intervals: $0\leq z<0.2$, $0.5\leq z<1.5$, and $1.5\leq z<2.5$. The $0\leq z<0.2$ sample is drawn from SDSS (Section \ref{sec:sdss}), while the higher-redshift samples are taken from the CANDELS survey 
(Sections \ref{sec:candels}). In total, our observational sample is composed of 416,550 SDSS galaxies 
and 78,051 CANDELS galaxies that have stellar masses above $10^{8} {\rm M_{\odot}}$. Their mass-SFR-size relations are presented in Section \ref{sec:results_obs}. We note that in order to facilitate a fair comparison with simulations, we adopt a lower stellar-mass cut of $10^{9.7} {\rm M_{\odot}}$ to further limit our observational sample when comparing to the simulations. This mass limit (applied to both simulations and observations) corresponds to $\lesssim 3000$ stellar particles per galaxy for both simulations, below which the simulated galaxies may suffer from finite resolution issue. Table \ref{tab:summary of samples>9.7 obs} summarizes the observational samples used in this study. 

\begin{table}[!ht]
    \flushleft
    \caption{Observational data used in this paper. 
    }
    \begin{tabular}{llll}
    \hline
        Survey & Redshift range & Number & Number  \\ 
          &  &  $\log\left(\frac{M_*}{M_{\odot}}\right)>8.0$ & $\log\left(\frac{M_*}{M_{\odot}}\right)>9.7$  \\
        \hline
        SDSS & $0<z<0.2$ &416,550 & 380,754 \\ 
        CANDELS & $0.5<z<1.5$ &40,209 & 6,522 \\ 
        CANDELS & $1.5<z<2.5$ &37,842 & 5,111 \\ \hline
    \end{tabular}
    \label{tab:summary of samples>9.7 obs}
\end{table}

\subsection{SDSS} 
\label{sec:sdss}
The galaxy sample for a nearby Universe (at $z<0.2$) is extracted from the Sloan Digital Sky Survey Data Release 7 \citep[SDSS DR7;][]{SDSSDR7}. With both photometry and spectroscopy, the survey provides us a comprehensive view of the low-redshift Universe.


For the SDSS galaxies in this work, the stellar masses are adopted from the MPA-JHU catalog, derived by fitting stellar population synthesis models to the photometry and spectral features \citep{Salim2017}. Star formation rates (SFR) are estimated from aperture-corrected H$\alpha$ emission-line luminosities, with dust attenuation corrections based on the Balmer decrement \citep{Brinchmann2004}. 
Measurements of galaxy size $R_{\rm e}$, defined as SDSS $r$-band half-light radius, are taken from the Upenn photometric catalog\footnote{http://alan-meert-website-aws.s3-website-us-east-1.amazonaws.com/fit\_catalog/download/index.html} \citep{Meert2015}.

In order to check whether there is any significant systematic difference between a central galaxy sample and the total galaxy sample, we identify central galaxies using the halo-based group catalog from \citet{YangXiaohu2007}, which designates the most massive member of each group as the central. 
Among the 416,550 SDSS galaxies in our sample, 296,557 are classified as central galaxies. 
We note that the difference in the mass-SFR-size relation between the two galaxy samples is marginal (see Fig.\,1). For this reason, we do not further distinguish between the central and satellite galaxy populations when making comparisons between simulations and observations but take the total galaxy sample for comparison within each redshift bin.

\subsection{CANDELS}  \label{sec:candels}
The Cosmic Assembly Near-infrared Deep Extragalactic Legacy Survey (CANDELS, \citet{candels_MAST}) is a landmark Multi-Cycle Treasury program with the \textit{Hubble Space Telescope} (\textit{HST}) designed to probe galaxy formation and evolution from the epoch of reionization through the peak of cosmic star formation ($z \sim 1.5-8$). The survey covers a total area of approximately 800~arcmin$^{2}$ across five well-studied extragalactic fields: GOODS-N, GOODS-S, EGS, UDS, and COSMOS \citep{Grogin2011, Koekemoer2011,Dahlen2013,Galametz2013,2013Guo,Santini2015,Stefanon2017,Nayyeri2017,Barro2019,Kodra2023}. 

In this work, we utilize galaxy properties derived from catalogs anchored by the \textit{HST}/WFC3 $F160W$ ($H$-band) selection, which provides a high degree of completeness for massive galaxies up to $z \sim 3$. The catalogs are taken from the High-Level Science Product (HLSP)\footnote{https://archive.stsci.edu/hlsp/candels} across five fields: GOODS-N (\citealt{Barro2019}), GOODS-S (\citealt{2013Guo}), UDS (\citealt{Galametz2013}), COSMOS (\citealt{Nayyeri2017}), and EGS (\citealt{Stefanon2017}).  
The primary galaxy properties used in this analysis are derived as follows:

\begin{itemize}
    \item \textbf{Redshifts:} We adopt the best-fit photometric redshifts ($z_{\rm best}$) from the CANDELS multi-method assembly catalogs \citep{Kodra2023}. These redshifts represent the optimized combination of multiple codes and have been validated against available spectroscopic samples.
    
    \item \textbf{Stellar Masses:} For the GOODS-S, UDS, COSMOS, and EGS fields, we utilize the median stellar masses ($M_{\rm med}$) from the multi-team SED fitting release of \citet{Santini2015}. For GOODS-N, where a multi-team median was not available, we adopt masses derived from \texttt{FAST} SED fitting \citep{Kirek2009FAST} as presented by \citet{Barro2019}. 
    
    \item \textbf{Star Formation Rates:} We utilize the ``SFR-ladder'' total star formation rates (${\rm SFR}_{\rm ladder, total}$) from the \citet{Barro2019} unified catalogs. These estimates combine UV and IR indicators (including \textit{Spitzer}/MIPS and \textit{Herschel}) to provide a robust census of both obscured and unobscured star formation.
    
    \item \textbf{Galaxy Sizes:} Morphological information is sourced from 2D surface brightness fitting of S\'{e}rsic profiles to the F160W imaging by \citet{vanDerWel2014}. The adopted size is the circularized half-light radius ($R_{\rm e}$) using the flux radius parameter ($\text{FLUX\_RADIUS}$).
    
\end{itemize}

\section{Simulation data} \label{sec:sim_data}

For the numerical simulations, we adopt the same three redshift bins as in observations and select snapshots from both EAGLE \citep
{Schaye2015, EAGLE-DR} and TNG100 \citep{Genel_2018MNRAS_TNG_SizeEvo, Nelson18_TNGcolor, Pillepich_et_al.(2018b), Springel_et_al.(2018), Marinacci18_TNGmagnetic, Naiman_et_al.(2018)} within these redshift ranges. Galaxies identified in the corresponding snapshots constitute the simulation samples to compare to the observational datasets. While we acknowledge the importance of quiescent populations, we have excluded galaxies with zero-SFR (in both simulations and observations) from the current analysis because our primary focus is the scaling relations and physical trends specific to galaxies in and around the star-forming main sequence. In practice, for both the observational and simulation samples, we only select galaxies with $-2 \leqslant \log(\mathrm{SFR}/[\rm M_{\odot}/yr]) \leqslant 3$. Below, we give a brief summary to the EAGLE and TNG100 simulations in Section \ref{sec:EAGLE} and \ref{sec:TNG}, respectively. 


\subsection{EAGLE} \label{sec:EAGLE}

EAGLE is a suite of cosmological hydrodynamical galaxy formation simulations (\citealt{Schaye2015, EAGLE-DR}) run with a modified \texttt{GADGET} code \citep{Springel2005_GADGET}. In this work, we use the \texttt{RefL0100N1504} simulation, which has a comoving box size of $(100/h~\mathrm{Mpc})^3$. The mass resolution is $1.81 \times 10^{6}~\mathrm{M_\odot}$ for baryonic matter and $9.7 \times 10^{6}~\mathrm{M_\odot}$ for dark matter, with a plummer-equivalent gravitational softening length of 0.7 kpc.  

The EAGLE simulations have been shown to reproduce key observed galaxy properties and scaling relations. \citet{Furong2015} demonstrated that EAGLE galaxies exhibit a bimodal distribution in specific star formation rate (sSFR), with the SFMS evolving consistently with observations despite a modest offset of 0.2–0.5 dex. Furthermore, \citet{Furong2017} examined the galaxy stellar mass--size relation, defining size as the stellar half-mass radius, and found that the simulations agree with observations to within $\sim 0.1$ dex at $z < 1$ and $0.2$--$0.3$ dex at $1 < z < 2$.

For this study, we take stellar mass, SFR and (three-dimensional) stellar half-mass radius $R_{\rm hsm}$ for each galaxy from the public data release\footnote{\url{https://icc.dur.ac.uk/Eagle/}} \citep{EAGLEteam2017,McAlpine2016}. In particular, the adopted mass, SFR and half mass radius were calculated within a 30 kpc aperture. For reliable measurements of galaxy size, we restrain our galaxy sample to only include galaxies that have stellar masses above $10^{9.7}M_{\odot}$, which corresponds to $\sim 3000$ stellar particles at the resolution level.

The final EAGLE sample consists of 55,736 galaxies from 14 snapshots across the three investigated redshift ranges.   
A summary of the sample is provided in Table~\ref{tab:summary of samples>9.7 sims}.




\subsection{TNG100} \label{sec:TNG}
The TNG100 run is part of the IllustrisTNG project, a suite of magnetohydrodynamical (MHD) cosmological simulations \citep{Genel_2018MNRAS_TNG_SizeEvo, Nelson18_TNGcolor, Pillepich_et_al.(2018b), Springel_et_al.(2018), Marinacci18_TNGmagnetic, Naiman_et_al.(2018)} performed with the moving-mesh code \texttt{AREPO} \citep{Springel2010}. 
The TNG100 simulation has a box of $(110.7\,\mathrm{Mpc})^3$ cosmological volume, 
a baryonic mass resolution of $1.4\times10^6~\mathrm{M_\odot}$ and a dark matter particle mass of $7.5\times10^6~\mathrm{M_\odot}$. Such resolution well matches that of \texttt{RefL0100N1504} from the EAGLE simulation suite.

Previous studies have demonstrated that TNG100 reproduces a broad range of observed galaxy properties and scaling relations, including the color bimodality, the stellar mass–size relation, and the SFMS \citep{Nelson18_TNGcolor, Pillepich_et_al.(2018b), Genel_2018MNRAS_TNG_SizeEvo}.  
Specifically, \citet{Donnari2018} demonstrated that it matches the SFMS at $z = 0$ with an intrinsic scatter of $\sim 0.3$ dex and reproduces the observed UVJ bimodality at $z < 2$. Using both three-dimensional stellar half-mass and two-dimensional half-light radii, \citet{Genel_2018MNRAS_TNG_SizeEvo} found that TNG100 galaxy sizes agree with observations to within $\sim 0.25$ dex over $0 \leq z \lesssim 2$, with quenched galaxies systematically smaller than star-forming systems \citep[also see][]{2019Huertas-Company}. 


For each TNG100 galaxy, the adopted stellar mass (\texttt{SubhaloMassType}) is defined as the total stellar mass bound to the subhalo, the effective radius $R_{\rm hsm}$ (\texttt{SubhaloHalfmassRadType}) is calculated as the three-dimensional stellar half-mass radius, and the SFR (\texttt{SubhaloSFR}) is computed as the total instantaneous SFR of all star-forming gas cells bound to the subhalo. All these properties were already calculated and provided by the TNG public catalog\footnote{\url{http://www.tng-project.org/data/}} \citep{Nelson2019a}. Again we only include galaxies that have stellar masses above $10^{9.7}M_{\odot}$ in our sample for comparison in order to guarantee a sufficiently large number of resolution elements. The final TNG100 sample comprises 45,251 galaxies from eight snapshots spreading over the three investigated redshift ranges (details see Table~\ref{tab:summary of samples>9.7 sims}).

\section{Calculating relative size of galaxies}
\label{sec:DRe}

In this section, we describe the method used to quantify the relative sizes of main-sequence galaxies with respect to systems of the same stellar mass and redshift for a given dataset. We emphasize that our analysis focuses on the relative sizes of both simulated and observed galaxies, rather than their absolute sizes. This approach largely removes the explicit dependence on stellar mass while retaining key information about variations in assembly history among galaxies of the same mass but with different levels of star formation activity (see \citealt{he2025symmetryfundamentalparametersgalaxies}).
Furthermore, when comparing the absolute sizes of galaxies from simulations to observations, various assumptions and corrections must be implemented to remove systematics. These biases arise from, for example, the differences between 3D and 2D projection measurements, the use of half-mass versus half-light radii, or other methodological discrepancies. By taking relative sizes for comparison, we minimize the impact and uncertainty rising from the adopted assumptions and corrections. We note that in terms of absolute sizes of galaxies, previous studies already revealed general agreement between the simulations and observations (e.g., \citealt{Furong2017, deGraaff2022_EAGLEmasssize, Genel_2018MNRAS_TNG_SizeEvo, 2019Huertas-Company}). In this study, we do not generate synthetic images for the simulated galaxies to account for the variety of observational effects. Instead, we directly adopt the cataloged half–stellar–mass radius, $R_{\rm hsm}$, as a proxy for the observed effective radius $R_{\rm e}$. We then use this quantity to compute the relative logarithmic size, $\Delta \log R_{\rm e}$, with respect to the one expected for the main-sequence galaxies at fixed stellar mass.

For each dataset from a given redshift range, we first determine the SFMS using an iterative fitting procedure. To do so, we fit a power law to the $M_*$--${\rm SFR}$ relation using all galaxies therein. We then exclude galaxies whose SFRs lie more than 1 dex below the fitted relation and refit the remaining sample. This process is repeated until no galaxies fall more than 1 dex below the best-fitting relation in a given iteration.
To further compute relative galaxy size $\Delta \log R_{\rm e}$, we additionally fit a power law to the stellar mass–size relation, $M_*$--$ R_{\rm e}$, for the above identified SFMS.
This power-law relation then defines the expected average size of SFMS galaxies of any given stellar mass $M_*$ and at a given redshift. 
The relative logarithmic size $\Delta \log R_{\rm e}$ is then calculated as the difference between the logarithmic size of a given galaxy and the expected logarithmic size according the fitted mass-size relation for galaxies on the SFMS. 



\begin{table} 
\flushleft
    \caption{Summary of the simulation data used in this paper. 
    }
    \begin{tabular}{lllll}
    \hline
        Simulation & Redshift range & Snapshot No. & Number   \\
             & &   &  $\log\left(\frac{M_*}{M_{\odot}}\right)>9.7$
        \\ \hline 
        TNG100 & $0<z<0.2$ & 099, 091, 084 & 17,796 \\ 
        TNG100 & $0.5<z<1.5$ & 067, 059, 050 & 18,716 \\ 
        TNG100 & $1.5<z<2.5$ & 040, 033 & 8,739 \\ \hline
        EAGLE & $0<z<0.2$ & 28, 27, 26 & 13,879 \\ 
        EAGLE & $0.5<z<1.5$ & 23, 22, 21, 20  & 32,444 \\ 
         & & 19, 18, 17 & \\
        EAGLE & $1.5<z<2.5$ & 16, 15, 14, 13 & 9,413 \\ \hline
    \end{tabular}

    \label{tab:summary of samples>9.7 sims}
\end{table}

\section{Results} \label{sec:results}


In this section, we first present the mass–size–SFR relation for the observational samples in Section~\ref{sec:results_obs}, and then compare the simulated galaxy populations with observations in Section~\ref{sec:candels}. For both analyses, the samples are divided into three redshift bins following \citet{Wuyts2011}: $0 \leq z < 0.2$, $0.5 \leq z < 1.5$, and $1.5 \leq z < 2.5$. 

\subsection{Observation} \label{sec:results_obs}

\begin{figure*}
\includegraphics[width=\linewidth]{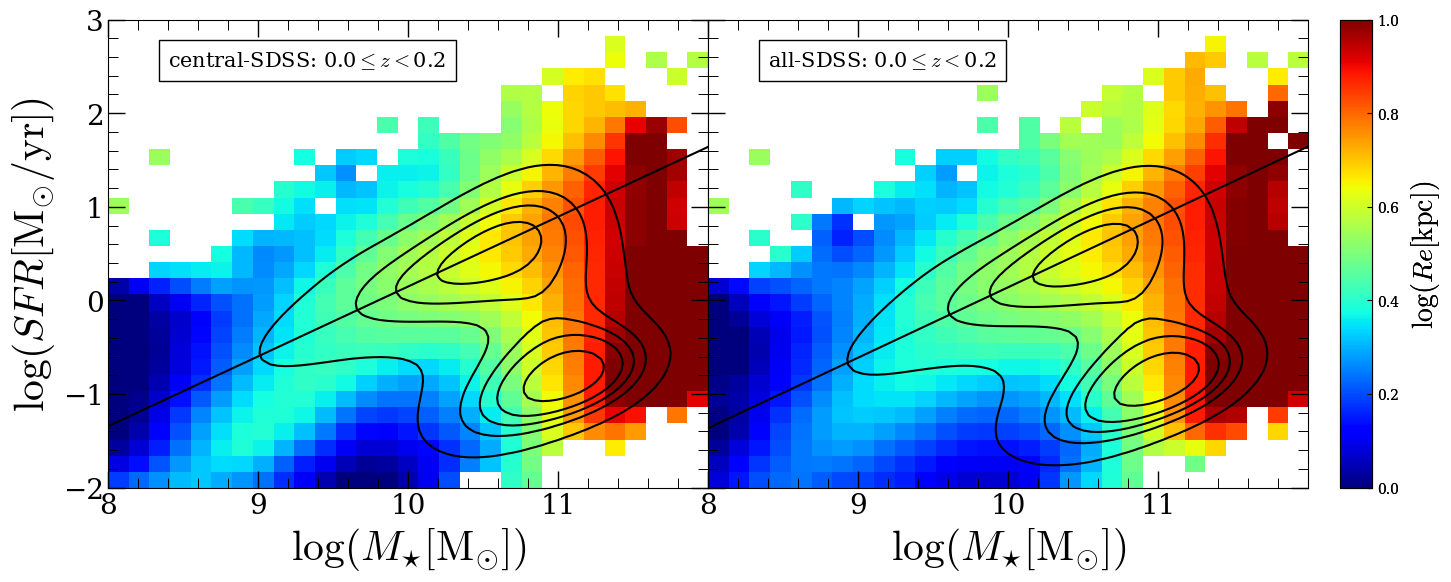}
\caption{
Mass–size–SFR relation of SDSS galaxies. The left panel shows central galaxies (296,557), and the right panel shows the full sample (416,550). Contours correspond to the 10th, 30th, 50th, 70th, and 90th percentiles of the joint distribution of stellar mass and star formation rate (SFR). The black line denotes the star-forming main sequence. The color scale represents the median galaxy size of galaxies within each pixel.
For clarity, the color field is smoothed using the gaussian\_filter function from the {\sc Scipy} package \citep{2020SciPy-NMeth}. The central galaxy sample is defined following \citet{YangXiaohu2007}.}
\label{fig:SDSS}
\end{figure*}

\begin{figure*}
\includegraphics[width=\linewidth]{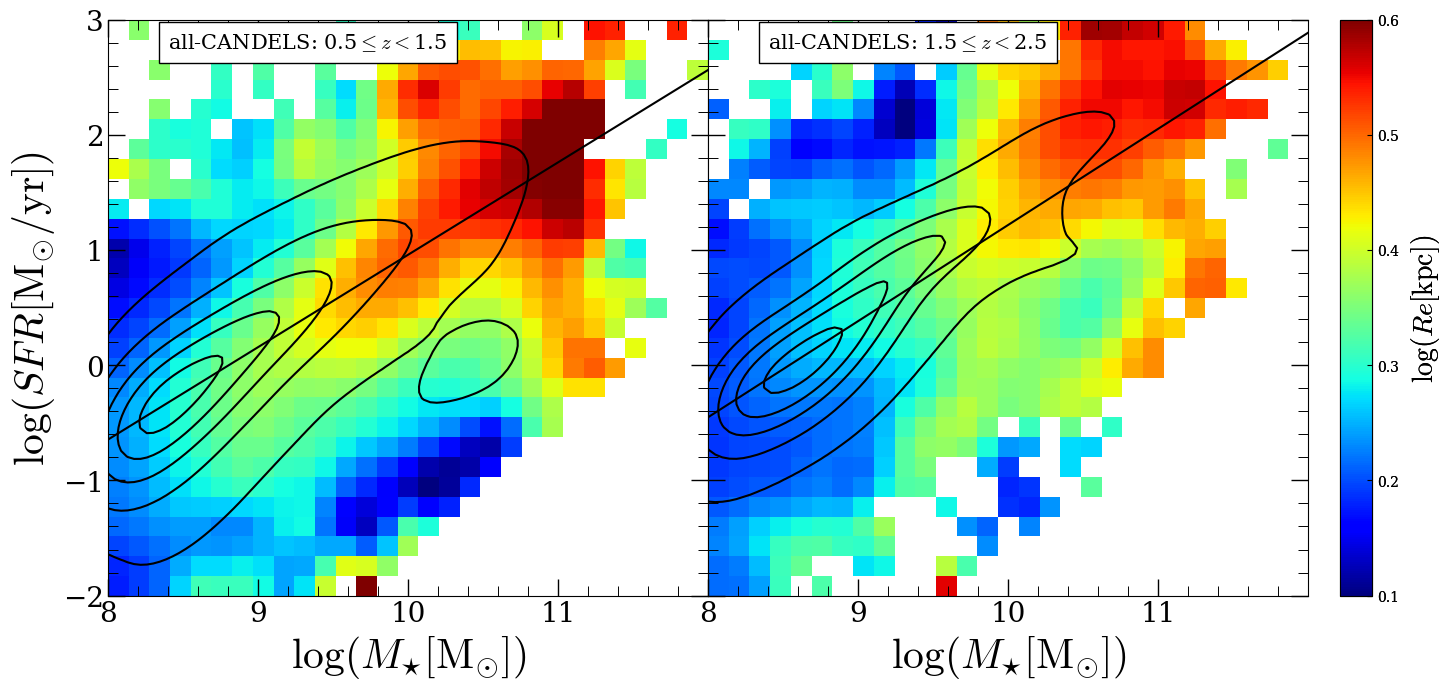}
\caption{Mass–size–SFR relation of CANDELS galaxies at $0.5<z<1.5$ (left, 40,209 galaxies) and $1.5<z<2.5$ (right, 37,842 galaxies), following the same plotting style as Fig.\,\ref{fig:SDSS}.
\label{fig:candels}}
\end{figure*}


To jointly examine the correlations among galaxy mass, size, and star formation rate, we color-code the median logarithmic galaxy size at fixed stellar mass and SFR in the $\log M_*-\log{\rm SFR}$ plane. For each galaxy sample in a given redshift bin, we also overplot the best-fit power-law relation defining the SFMS, together with galaxy number-density contours that enclose 10\%, 30\%, 50\%, 70\% and 90\% of the sample. In all cases, the fitted power-law closely traces the elongation of the density contours, indicating that it effectively captures the ridge of the SFMS where the majority of galaxies reside.

Fig.\,\ref{fig:SDSS} shows the results for all SDSS galaxies with $\log (M_*/M_{\odot})\geqslant 8$ at $z<0.2$, central galaxies (see Section \ref{sec:sdss} for the method of their identification) on the left and total galaxy population on the right. As can be seen, the two samples exhibit very similar overall trends, suggesting that environmental effects do not play a primary role in shaping the joint mass–SFR–size relation. In general, more massive galaxies exhibit larger sizes, as already established in many previous works \citep{Daddi2005, vanDokkum2008, Newman2012, Barro2013, Carollo2013}. At a fixed stellar mass above $\log M_*/M_{\odot} \sim 9.5$, galaxies living around the ridge of the SFMS possess larger sizes than galaxies living further away. At lower masses, the largest galaxies from each of the stellar mass bins collectively follow a somewhat steeper $\log M_* - \log {\rm SFR}$ relation, i.e., having lower SFR than expected from the best-fit power-law relation. We note that the overall pattern as well as this particular trend at lower masses are also evident and all consistent with \citet{Wuyts2011}. Regarding this trend at lower masses, it may be driven by larger aperture corrections for extended galaxies, which can potentially introduce systematic biases. Notably, \citet{he2025symmetryfundamentalparametersgalaxies} carried out a comparable analysis based on SDSS data, using SFR measurements inferred from spectral energy distribution fitting \citep{Salim2017}, and reported no such features at the low-mass end.
 

Fig.\,\ref{fig:candels} shows the results for all CANDELS galaxies (regardless of their central or satellite status) split into $0.5\leqslant z<1.5$ (left) and $1.5\leqslant z<2.5$ (right) redshift bins. Once again, the trend that galaxies on the SFMS ridge are systematically larger than those lying farther from it—well established in the local Universe—remains prominent at $0.5 \leq z < 1.5$, but becomes weaker and is no longer clearly evident at $1.5 \leq z < 2.5$. However, SFR (and other) measurements at higher redshift suffer from increased uncertainties, primarily due to the greater luminosity distances involved. 
We conclude that the observed mass–size–SFR relation is well established at least up to $z\sim 1.5$: at fixed stellar mass, galaxy sizes decrease away from the ridge of the SFMS toward both higher and lower SFRs.

\subsection{Simulations vs. Observations} 
\label{sec:obsvssims}

\begin{figure*}
\includegraphics[width=\linewidth]{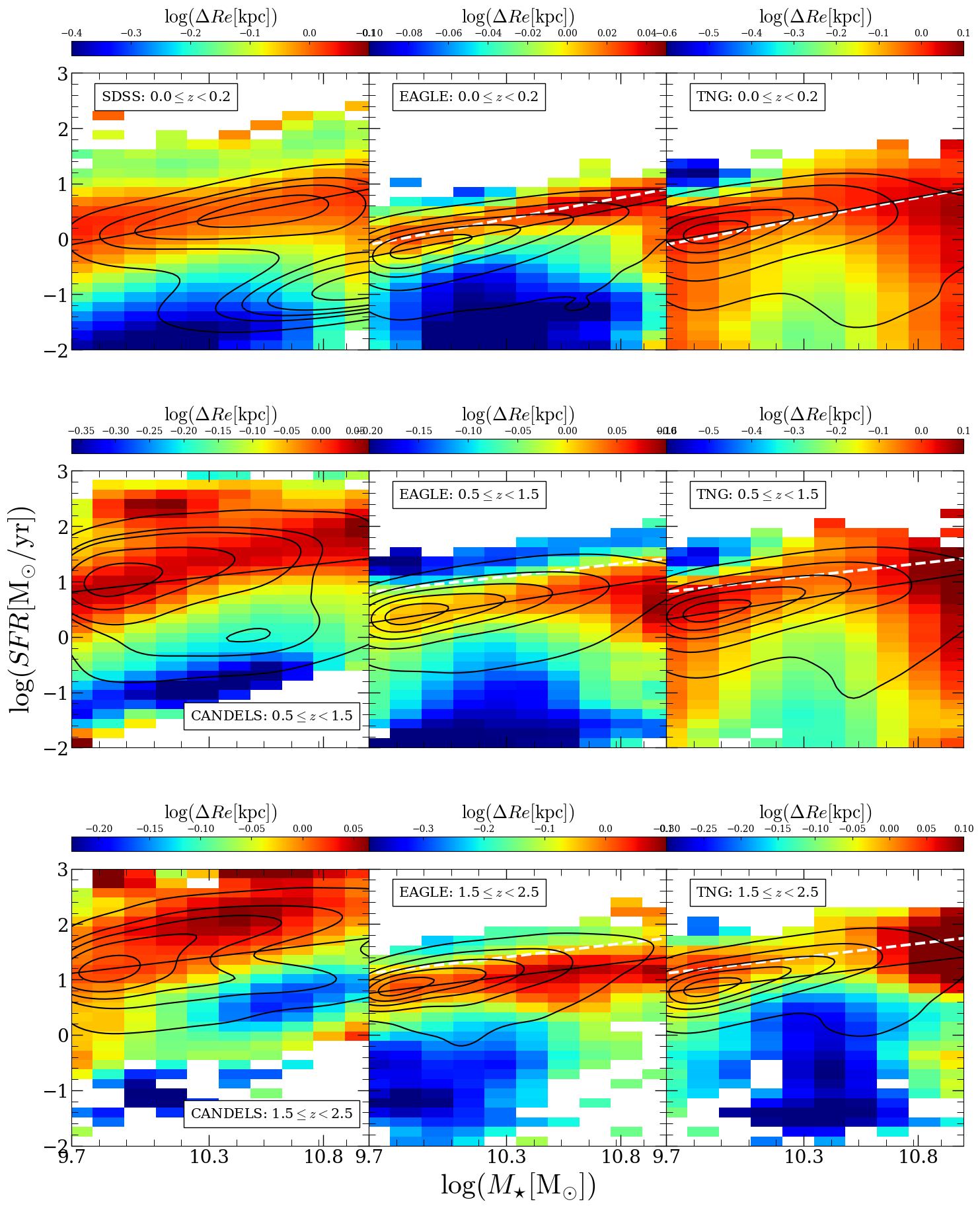}
\caption{Relative galaxy size, $\Delta \log{\rm SFR}$, as a function of galaxy SFR and stellar mass: panels are arranged from top to bottom in order of increasing redshift, and from left to right for observations, EAGLE, and TNG100, as indicated in the legend,
same plotting style as in Fig.,\ref{fig:SDSS}. In the simulation panels, the corresponding observational SFMS relation is overlaid as a white dashed line. Note that only galaxies with stellar masses greater than $10^{9.7},\rm M_{\odot}$ (above the stellar mass completeness limit of both blue and red galaxies out to z=2.5, see \citealt{Tomczak2014}) are included in this comparison.  
\label{fig:pixeldRe}}
\end{figure*}


\begin{figure*}
\includegraphics[width=\linewidth]{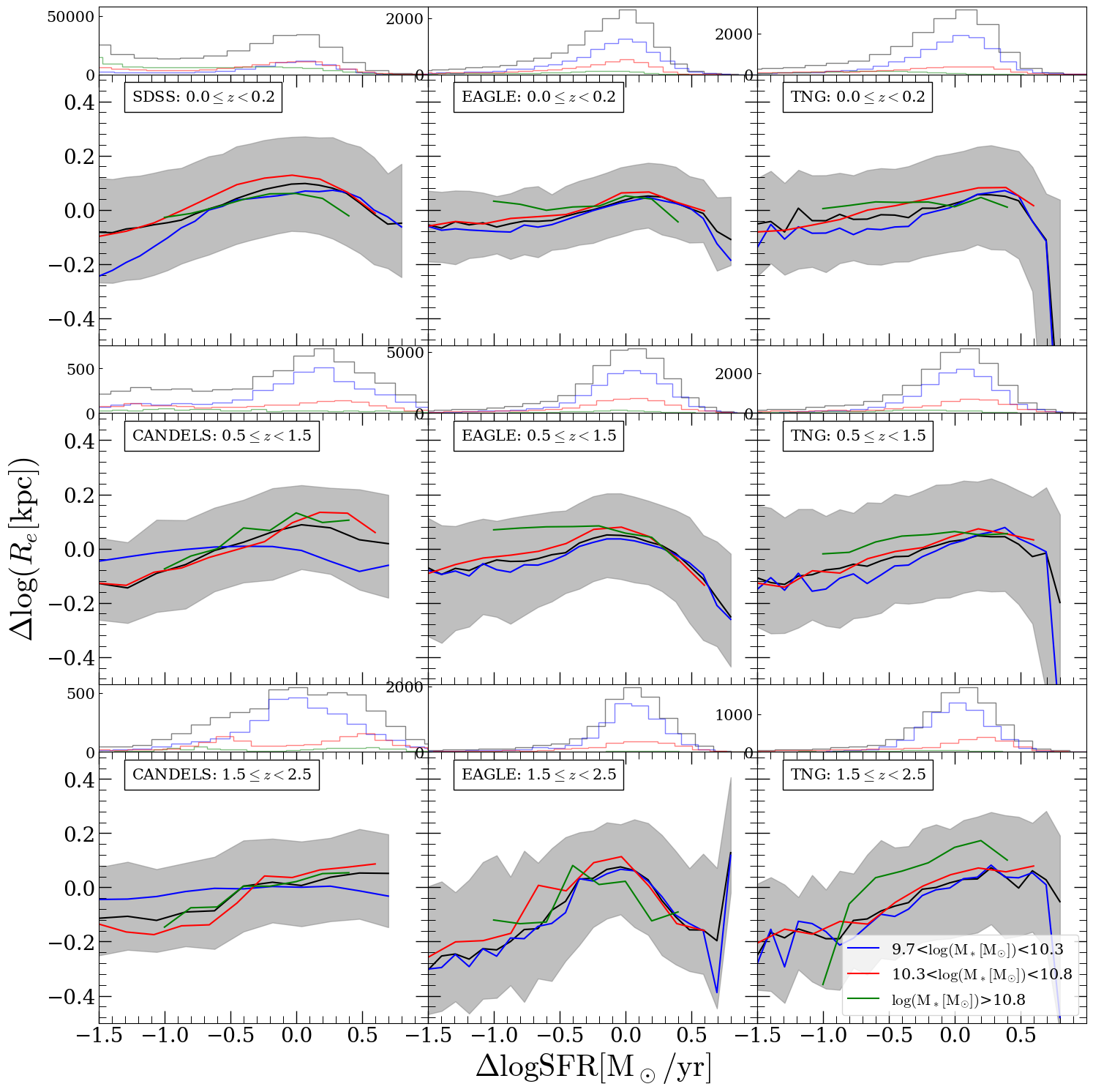}
\caption{Relative galaxy size, $\Delta \log R_{\rm e}$, as a function of star formation rate offset, $\Delta \log \mathrm{SFR}$ (i.e., deviation from the ridge of SFMS). Panels are arranged from top to bottom in order of increasing redshift, and from left to right for the observational sample, EAGLE, and TNG100, as indicated in the legend. In each panel, the black solid line shows the median $\Delta \log R_{\rm e}$ as a function of $\Delta \log \mathrm{SFR}$, while the gray shaded region encloses the 16th--84th percentiles. The blue, red, and green lines represent the median relations for low-mass, intermediate-mass, and massive galaxies, respectively. Histograms above each panel show the distribution of $\Delta \log \mathrm{SFR}$. }
\label{fig:dRe}
\end{figure*}



In this section, we compare simulated galaxies with observations. We restrict the simulated sample to galaxies with stellar masses $\log (M_*/M_{\odot}) \geq 9.7$, corresponding to systems resolved with at least several thousand stellar particles, ensuring that galaxy structures and sizes are reliably measured. For consistency, we apply the same stellar-mass cut to the observational sample.


As discussed above, our analysis focuses on the relative galaxy size, $\Delta \log R_{\rm e}$, defined as the logarithmic offset between a galaxy’s size and the expected size at fixed stellar mass predicted by the best-fit mass–size relation for SFMS galaxies (see Section~\ref{sec:DRe}). This relative metric removes the primary mass dependence, thereby isolating variations associated with differences in assembly history and star formation activity. It also mitigates systematic uncertainties when comparing simulations with observations.


Figure \ref{fig:pixeldRe} presents the relative size, $\Delta \log R_{\rm e}$, color-coded in the $\log M_*-\log{\rm SFR}$ plane. From left to right, the three columns show results for the observational sample, the EAGLE simulation, and the TNG100 simulation, respectively, while the three rows correspond to different redshift bins. 
As can be seen, the left panels are essentially the same as results presented in Fig. \ref{fig:SDSS} and Fig. \ref{fig:candels} for the observed galaxies, only here in terms of their relative sizes and restricted to the adopted stellar mass range. For the simulated galaxies in the lowest redshift bin ($z<0.2$), normalizations of SFMS (as indicated by the lines from the best-fit power law models) of both EAGLE and TNG100 simulations are generally consistent with that of the observed galaxies. However, at higher redshifts the normalizations from both simulations are systematically lower than the observed galaxies. These offsets have already been reported by previous studies: the normalizations at these higher redshifts are typically 0.2-0.5 dex below observational constraints, associated also with relatively lower gas fractions in simulated galaxies at the cosmic noon (see \citealt{Furong2015, Donnari2018}).


When considering relative size, the two simulations show broadly similar color-coded distributions, although their detailed patterns differ noticeably from the observations. Nevertheless, both simulations reproduce the observed trend that galaxies on the SFMS ridge are systematically larger than those with higher or lower SFRs. This feature is present across the full redshift range and over a wide stellar-mass range explored in this study, indicating that it is robust and likely reflects general physical processes rather than being driven by specific processes or transient effects. We discuss a possible physical origin of this trend in Section \ref{sec:conclusion}.

To illustrate this behavior more clearly, Fig.~\ref{fig:dRe} shows the relative size, $\Delta \log R_{\rm e}$, as a function of the main-sequence offset, $\Delta \log{\rm SFR} \equiv \log({\rm SFR}) - \log({\rm SFR}_{\rm SFMS})$, where ${\rm SFR}_{\rm SFMS}$ is given by the best-fit SFMS. For each observational and simulated sample, offsets from the star-forming main sequence are computed relative to the best-fitting main sequence derived from that sample itself. The blue, red, and green curves indicate the median relations for galaxies with $9.7 < \log(M_*/M_\odot) < 10.3$, $10.3 < \log(M_*/M_\odot) < 10.8$, and $\log(M_*/M_\odot) > 10.8$, respectively. As can be seen, the relative size of galaxies generally decreases as they move away from the SFMS ridge, although the strength of this trend varies with redshift and between observations and simulations. In the observational sample, the trend is particularly evident in the two lower-redshift bins ($z < 1.5$) but becomes insignificant at $1.5 < z < 2.5$.

In contrast, the EAGLE simulation exhibits this trend across all redshifts, with a pronounced strengthening toward higher redshift. Particularly at $z > 0.5$, the relative size $\Delta \log R_{\rm e}$ of the most starbursting galaxies ($\Delta \log{\rm SFR} > 0.5$) declines more steeply than in the observations, implying that high-redshift starbursts in EAGLE are systematically more compact at fixed stellar mass and SFR than seen in observations. The TNG100 simulation on the other hand shows the weakest overall dependence of the relative size $\Delta \log R_{\rm e}$ on the SFMS offset $\Delta \log{\rm SFR}$, both compared to EAGLE and to the observations. At $z > 0.5$, the two simulations diverge notably: starburst galaxies in TNG100 are significantly less compact than their EAGLE counterparts.


Finally, we also note that the above-mentioned galaxy mass-SFR-size relation and the $\Delta \log R_{\rm e}-\Delta \log{\rm SFR}$ relation are essentially dominated by galaxies with $\log(M_*/M_\odot) < 10.8$ (blue and red curves), while the most massive galaxies (green curves) exhibit little or no systematic trend. 

\section{Summary and Discussion} \label{sec:conclusion}

The observed mass–SFR–size relation of galaxies up to $z\sim 2.5$ reveals a deep connection between the evolution of galaxy structure/morphology and stellar populations, both regulated by galaxy assembly history and star-formation history. In this study, we compare the mass–SFR–size relations of galaxies from both EAGLE and TNG100 simulations to those from SDSS and CANDELS galaxies across three redshift bins ($0\leq z<0.2$, $0.5\leq z<1.5$, and $1.5\leq z<2.5$). We note that the observational sizes are projected half-light radii, whereas the simulation sizes are intrinsic half-mass radii. In addition, the EAGLE and TNG catalogs adopt slightly different operational size definitions. Therefore, the comparison should primarily be interpreted in a relative rather than absolute sense. For this reason, we focus on comparisons in terms of relative galaxy sizes $\Delta \log R_{\rm e}$ measured with respect to the average size of SFMS galaxies at fixed stellar mass in order to eliminate the dependence of galaxy size on stellar mass (see Section \ref{sec:DRe} for definition). 

The observational data exhibit a discernible trend wherein galaxy size decreases with increasing offset away from the main-sequence ridge at redshifts $z < 1.5$, i.e., galaxies living above and below the ridge tend to be systematically more compact than those around the ridge, while at redshifts $z > 1.5$, such a trend is no longer detectable (see Section\,\ref{sec:results_obs} and Figs.\,1 and 2). Such behaviors are in good agreement with the original discovery reported in \citet{Wuyts2011}. In comparison, the same trend is generally present in both EAGLE and TNG100 simulations and across all three redshift bins. However, it is more pronounced in the former simulation than in the latter. Such a trend is largely dominated by galaxies with stellar masses below $\log(M_*/M_{\odot})\sim 10.8$. For the most starbursting galaxies (i.e., with $\Delta \log \rm{SFR}>0.5$) at redshifts $z>0.5$, the relative sizes $\Delta \log R_{\rm e}$ of galaxies decrease much faster towards higher $\Delta \log \rm{SFR}$ in the EAGLE simulation than for the TNG100 galaxies and than in the observations (see Section \ref{sec:obsvssims} and Figs.\,3 and 4). 

We note that a more rigorous comparison between the simulations and observations would require using realistic mock observations from forward-modeling, including for example, dust attenuation, mock photometry, observational apertures, and re-derived stellar masses, SFRs, and half-light radii etc. Previous studies have shown that such procedures can alter inferred galaxy properties quantitatively (e.g., \citealt{Torrey2015, RodriguezGomez2019, Snyder2015}), although the qualitative trends often remain robust. Illustrating that this is not always the case, \citet{Katsianis2020} applied different observational SFR derivation techniques to realistic mock observations generated with SKIRT for EAGLE galaxies and showed that the tension between the simulation and observations (as well as among different observations) can be largely alleviated when methods are matched consistently. 

Many physical processes may affect SFRs and shape the size growth of galaxies (see introduction). Modern cosmological simulations routinely utilize subgrid models to capture these processes, though their specific implementation varies across different simulations. It is worth noting that neither EAGLE nor TNG100 simulation were designed or calibrated to reproduce this joint mass–SFR–size relation. However, both simulations are able to largely reproduce the observed trend in the mass-SFR-size relation. This reflects that it more likely arises from the standard framework commonly adopted by modern cosmological simulations, in combination with the inherent gravitational physics that N-body methods naturally encode. 


One possible explanation of the observed mass-SFR-size relation comes from time variability in star formation of SFMS galaxies. In this regard, many observations have revealed possible diagnostics of fluctuating star formation histories, in particular among lower mass galaxies (\citealt{Tolstoy2009ARAA, McQuinn2010a, McQuinn2010b, Weisz2011, EnciWang_and_Lilly2020a, JunYin2023, 2025MunozLopez, MarissaPerry2025, JennyWan2025}). Correspondingly, simulations have also predicted such fluctuating behaviors of galaxies across a wide range of redshifts (\citealt{Quenching_Zolotov2015, SFGGas_Tacchella2016, Tacchella2020, Cenci2024, Fortune2025, THESANZOOM_McClymont2025}). In particular, recent studies by \citet{SenWang_2022} and \citet{Gui2025} carried out detailed investigations in this regard for SFMS galaxies at $z<1$ in the TNG100 simulation. Within the framework of cosmological simulations, the temporal fluctuation in SFR is a natural result of cold gas replenishment and depletion modulated by both internal feedback processes (i.e., feedback-driven breathing cycles) and the ambient angular momentum environment (see \citealt{WangSen2025}). Galaxies thus experience a rise and fall in star formation, synchronized with rhythmic variations in cold gas content at circumgalactic scales. Such a temporal fluctuation in SFR can also explain $\sim$80\% of the variance in the present-day SFMS (\citealt{Gui2025}) and the observed fundamental metallicity relation (e.g., \citealt{Mannucci2010}, see Mei et al. in prep).

In contrast, galaxy size, however, is an integrated property that resists variation on shorter timescales. We shall note that at shorter, rest-UV wavelengths sizes may indeed be modulated more rapidly in the presence of star formation variations across galaxy disks. However, in the case where the sizes are quantified on the basis of the accumulated stellar mass distributions, or half-light radii measured at rest-frame longer wavelengths, these size indicators serve as a good proxy of the stellar mass distribution. As such, significant evolution in galaxy size typically requires disruptive processes such as major mergers. Therefore if we consider galaxy size as a more stable and even fundamental property to distinguish galaxies among one another, one can interpret the observed joint mass-SFR-size relation through the dependence of the temporal variation in SFR on galaxy size: At fixed stellar mass, more compact galaxies exhibit a larger temporal scatter in SFR than their more extended galaxy counterparts, while having comparable time-averaged SFRs—thus, both more compact and more extended galaxies share the same SFMS ridge. 

Such an explanation has already been postulated and tested to a certain extent using observational data and through analytical modeling. 
In \citet{Wang2019}, the authors suggested that as compact star-forming galaxies tend to possess lower neutral hydrogen gas content (\citealt{Wang2018}) and thus are expected to have higher star formation efficiency compared to their more extended counterparts \citep[e.g.,][]{Shi2011}, they are able to convert cold gas into stars more efficiently. This makes them experience more bursty star formation for a given amount of gas accretion, placing them on the upper envelope of the SFMS. They then consume their cold gas more rapidly than more extended galaxies and subsequently migrate toward the lower envelope of the SFMS. Such an evolutionary sequence thus naturally explains the pronounced mass–SFR–size relation identified in observations (\citealt{Wuyts2011, he2025symmetryfundamentalparametersgalaxies}). Building upon this idea, 
the authors developed a gas-regulator model \citep{Lilly2013} in which galaxy evolution is driven by time variations in the gas inflow rate \citep[see also, e.g.,][]{Tacchella2020, SenWang_2022, Gui2025}. Utilizing the spatially resolved spectroscopic data of MaNGA galaxies, the model quantitatively reproduced the observed scatter in SFR both within individual galaxies and across the galaxy population. In addition, the model also successfully accounted for the observed anti-correlation between SFR and gas-phase metallicity seen in both observations and simulations \citep{Mannucci2010, Wang2021, Ma2024}.


Our study demonstrates that both simulations broadly reproduce the observed trend that galaxy size decreases with increasing offset away from the SFMS. It is worth noting that these properties actually emerge self-consistently from the simulation, without deliberately or simultaneously calibrating the mass–size–SFR relation to match the observational trends. In light of this, it is highly interesting to use cosmological simulations to examine the validity of such a postulation, where more compact galaxies have larger temporal fluctuations in SFR (as a consequence of higher star-formation efficiency) and thus are more prevalent among the upper and lower envelopes of the SFMS, in comparison to their more extended counterparts. 

\section*{Acknowledgments}
We thank Yuqian Gui and Drs. Lizhi Xie, Lan Wang for useful discussion. We also thank an anonymous referee for useful and constructive comments. This work is supported by the National Key Research and Development Program of China (Grant No. 2022YFA1602903), the National Natural Science Foundation of China (Grant No. 12433003 and 12473008), and the China Manned Space Project (No. CMS-CSST-2025-A10). The authors gratefully acknowledge support from the Royal Society International Exchanges scheme (IES\textbackslash R2\textbackslash 242195). SW acknowledges support from China’s National Foreign Expert programme (H20240871). This work acknowledges the Tsinghua Astrophysics High-Performance Computing platform for providing the computational and storage resources that supported this research. 


\bibliography{reference}

@ARTICLE{2020SciPy-NMeth,
  author  = {Virtanen, Pauli and Gommers, Ralf and Oliphant, Travis E. and
            Haberland, Matt and Reddy, Tyler and Cournapeau, David and
            Burovski, Evgeni and Peterson, Pearu and Weckesser, Warren and
            Bright, Jonathan and {van der Walt}, St{\'e}fan J. and
            Brett, Matthew and Wilson, Joshua and Millman, K. Jarrod and
            Mayorov, Nikolay and Nelson, Andrew R. J. and Jones, Eric and
            Kern, Robert and Larson, Eric and Carey, C J and
            Polat, {\.I}lhan and Feng, Yu and Moore, Eric W. and
            {VanderPlas}, Jake and Laxalde, Denis and Perktold, Josef and
            Cimrman, Robert and Henriksen, Ian and Quintero, E. A. and
            Harris, Charles R. and Archibald, Anne M. and
            Ribeiro, Ant{\^o}nio H. and Pedregosa, Fabian and
            {van Mulbregt}, Paul and {SciPy 1.0 Contributors}},
  title   = {{{SciPy} 1.0: Fundamental Algorithms for Scientific
            Computing in Python}},
  journal = {Nature Methods},
  year    = {2020},
  volume  = {17},
  pages   = {261--272},
  adsurl  = {https://rdcu.be/b08Wh},
  doi     = {10.1038/s41592-019-0686-2},
}

@ARTICLE{Planck_Collaboration2016,
       author = {{Planck Collaboration} and {Ade}, P.~A.~R. and {Aghanim}, N. and {Arnaud}, M. and {Ashdown}, M. and {Aumont}, J. and {Baccigalupi}, C. and {Banday}, A.~J. and {Barreiro}, R.~B. and {Bartlett}, J.~G. and {Bartolo}, N. and {Battaner}, E. and {Battye}, R. and {Benabed}, K. and {Beno{\^\i}t}, A. and {Benoit-L{\'e}vy}, A. and {Bernard}, J. -P. and {Bersanelli}, M. and {Bielewicz}, P. and {Bock}, J.~J. and {Bonaldi}, A. and {Bonavera}, L. and {Bond}, J.~R. and {Borrill}, J. and {Bouchet}, F.~R. and {Boulanger}, F. and {Bucher}, M. and {Burigana}, C. and {Butler}, R.~C. and {Calabrese}, E. and {Cardoso}, J. -F. and {Catalano}, A. and {Challinor}, A. and {Chamballu}, A. and {Chary}, R. -R. and {Chiang}, H.~C. and {Chluba}, J. and {Christensen}, P.~R. and {Church}, S. and {Clements}, D.~L. and {Colombi}, S. and {Colombo}, L.~P.~L. and {Combet}, C. and {Coulais}, A. and {Crill}, B.~P. and {Curto}, A. and {Cuttaia}, F. and {Danese}, L. and {Davies}, R.~D. and {Davis}, R.~J. and {de Bernardis}, P. and {de Rosa}, A. and {de Zotti}, G. and {Delabrouille}, J. and {D{\'e}sert}, F. -X. and {Di Valentino}, E. and {Dickinson}, C. and {Diego}, J.~M. and {Dolag}, K. and {Dole}, H. and {Donzelli}, S. and {Dor{\'e}}, O. and {Douspis}, M. and {Ducout}, A. and {Dunkley}, J. and {Dupac}, X. and {Efstathiou}, G. and {Elsner}, F. and {En{\ss}lin}, T.~A. and {Eriksen}, H.~K. and {Farhang}, M. and {Fergusson}, J. and {Finelli}, F. and {Forni}, O. and {Frailis}, M. and {Fraisse}, A.~A. and {Franceschi}, E. and {Frejsel}, A. and {Galeotta}, S. and {Galli}, S. and {Ganga}, K. and {Gauthier}, C. and {Gerbino}, M. and {Ghosh}, T. and {Giard}, M. and {Giraud-H{\'e}raud}, Y. and {Giusarma}, E. and {Gjerl{\o}w}, E. and {Gonz{\'a}lez-Nuevo}, J. and {G{\'o}rski}, K.~M. and {Gratton}, S. and {Gregorio}, A. and {Gruppuso}, A. and {Gudmundsson}, J.~E. and {Hamann}, J. and {Hansen}, F.~K. and {Hanson}, D. and {Harrison}, D.~L. and {Helou}, G. and {Henrot-Versill{\'e}}, S. and {Hern{\'a}ndez-Monteagudo}, C. and {Herranz}, D. and {Hildebrandt}, S.~R. and {Hivon}, E. and {Hobson}, M. and {Holmes}, W.~A. and {Hornstrup}, A. and {Hovest}, W. and {Huang}, Z. and {Huffenberger}, K.~M. and {Hurier}, G. and {Jaffe}, A.~H. and {Jaffe}, T.~R. and {Jones}, W.~C. and {Juvela}, M. and {Keih{\"a}nen}, E. and {Keskitalo}, R. and {Kisner}, T.~S. and {Kneissl}, R. and {Knoche}, J. and {Knox}, L. and {Kunz}, M. and {Kurki-Suonio}, H. and {Lagache}, G. and {L{\"a}hteenm{\"a}ki}, A. and {Lamarre}, J. -M. and {Lasenby}, A. and {Lattanzi}, M. and {Lawrence}, C.~R. and {Leahy}, J.~P. and {Leonardi}, R. and {Lesgourgues}, J. and {Levrier}, F. and {Lewis}, A. and {Liguori}, M. and {Lilje}, P.~B. and {Linden-V{\o}rnle}, M. and {L{\'o}pez-Caniego}, M. and {Lubin}, P.~M. and {Mac{\'\i}as-P{\'e}rez}, J.~F. and {Maggio}, G. and {Maino}, D. and {Mandolesi}, N. and {Mangilli}, A. and {Marchini}, A. and {Maris}, M. and {Martin}, P.~G. and {Martinelli}, M. and {Mart{\'\i}nez-Gonz{\'a}lez}, E. and {Masi}, S. and {Matarrese}, S. and {McGehee}, P. and {Meinhold}, P.~R. and {Melchiorri}, A. and {Melin}, J. -B. and {Mendes}, L. and {Mennella}, A. and {Migliaccio}, M. and {Millea}, M. and {Mitra}, S. and {Miville-Desch{\^e}nes}, M. -A. and {Moneti}, A. and {Montier}, L. and {Morgante}, G. and {Mortlock}, D. and {Moss}, A. and {Munshi}, D. and {Murphy}, J.~A. and {Naselsky}, P. and {Nati}, F. and {Natoli}, P. and {Netterfield}, C.~B. and {N{\o}rgaard-Nielsen}, H.~U. and {Noviello}, F. and {Novikov}, D. and {Novikov}, I. and {Oxborrow}, C.~A. and {Paci}, F. and {Pagano}, L. and {Pajot}, F. and {Paladini}, R. and {Paoletti}, D. and {Partridge}, B. and {Pasian}, F. and {Patanchon}, G. and {Pearson}, T.~J. and {Perdereau}, O. and {Perotto}, L. and {Perrotta}, F. and {Pettorino}, V. and {Piacentini}, F. and {Piat}, M. and {Pierpaoli}, E. and {Pietrobon}, D. and {Plaszczynski}, S. and {Pointecouteau}, E. and {Polenta}, G. and {Popa}, L. and {Pratt}, G.~W. and {Pr{\'e}zeau}, G.},
        title = "{Planck 2015 results. XIII. Cosmological parameters}",
      journal = {\aap},
         year = 2016,
        month = sep,
       volume = {594},
          eid = {A13},
        pages = {A13},
          doi = {10.1051/0004-6361/201525830},
archivePrefix = {arXiv},
       eprint = {1502.01589},
 primaryClass = {astro-ph.CO},
       adsurl = {https://ui.adsabs.harvard.edu/abs/2016A&A...594A..13P}
}

@ARTICLE{Martorano2026,
       author = {{Martorano}, M. and {van der Wel}, A. and {Gebek}, A. and {Baes}, M. and {Bell}, E.~F. and {Brammer}, G. and {Meidt}, S.~E. and {Nersesian}, A. and {Whitaker}, K. and {Wuyts}, S.},
        title = "{Evolution and mass dependence of UV-to-near-IR color gradients up to z = 2.5 from the Hubble Space Telescope and the James Webb Space Telescope}",
      journal = {\aap},
         year = 2026,
        month = jan,
       volume = {705},
          eid = {A236},
        pages = {A236},
          doi = {10.1051/0004-6361/202555974},
archivePrefix = {arXiv},
       eprint = {2512.02440},
 primaryClass = {astro-ph.GA},
       adsurl = {https://ui.adsabs.harvard.edu/abs/2026A&A...705A.236M}
}

@ARTICLE{vanderWel2024,
       author = {{van der Wel}, Arjen and {Martorano}, Marco and {H{\"a}u{\ss}ler}, Boris and {Nedkova}, Kalina V. and {Miller}, Tim B. and {Brammer}, Gabriel B. and {van de Ven}, Glenn and {Leja}, Joel and {Bezanson}, Rachel S. and {Muzzin}, Adam and {Marchesini}, Danilo and {de Graaff}, Anna and {Nelson}, Erica J. and {Kriek}, Mariska and {Bell}, Eric F. and {Franx}, Marijn},
        title = "{Stellar Half-mass Radii of 0.5 z < 2.3 Galaxies: Comparison with JWST/NIRCam Half-light Radii}",
      journal = {\apj},
         year = 2024,
        month = jan,
       volume = {960},
       number = {1},
          eid = {53},
        pages = {53},
          doi = {10.3847/1538-4357/ad02ee},
archivePrefix = {arXiv},
       eprint = {2307.03264},
 primaryClass = {astro-ph.GA},
       adsurl = {https://ui.adsabs.harvard.edu/abs/2024ApJ...960...53V}
}

@ARTICLE{SenWang_2022,
       author = {{Wang}, Sen and {Xu}, Dandan and {Lu}, Shengdong and {Cai}, Zheng and {Xiang}, Maosheng and {Mao}, Shude and {Springel}, Volker and {Hernquist}, Lars},
        title = "{From large-scale environment to CGM angular momentum to star-forming activities - I. Star-forming galaxies}",
      journal = {\mnras},
         year = 2022,
        month = jan,
       volume = {509},
       number = {3},
        pages = {3148-3162},
          doi = {10.1093/mnras/stab3167},
archivePrefix = {arXiv},
       eprint = {2109.06200},
 primaryClass = {astro-ph.GA},
       adsurl = {https://ui.adsabs.harvard.edu/abs/2022MNRAS.509.3148W}
}

@ARTICLE{THESANZOOM_McClymont2025,
       author = {{McClymont}, William and {Tacchella}, Sandro and {Smith}, Aaron and {Kannan}, Rahul and {Puchwein}, Ewald and {Borrow}, Josh and {Garaldi}, Enrico and {Keating}, Laura and {Vogelsberger}, Mark and {Zier}, Oliver and {Shen}, Xuejian and {Popovic}, Filip and {Simmonds}, Charlotte},
        title = "{The THESAN-ZOOM project: Burst, quench, repeat -- unveiling the evolution of high-redshift galaxies along the star-forming main sequence}",
      journal = {arXiv e-prints},
         year = 2025,
        month = feb,
          eid = {arXiv:2503.00106},
        pages = {arXiv:2503.00106},
          doi = {10.48550/arXiv.2503.00106},
archivePrefix = {arXiv},
       eprint = {2503.00106},
 primaryClass = {astro-ph.GA},
       adsurl = {https://ui.adsabs.harvard.edu/abs/2025arXiv250300106M}
}

@ARTICLE{Fortune2025,
       author = {{Fortun\'e}, Silvio amnd {Remus}, Rhea-Silvia and {Kimmig}, Lucassor C. and {Burkert}, Andreas and {Dolag}, Klaus},
        title = "{Die Hard: The On-Off-Cycle of Galaxies on the Star Formation Main Sequence}",
      journal = {arXiv e-prints},
         year = 2025,
        month = mar,
          eid = {arXiv:2503.20858},
        pages = {arXiv:2503.20858},
          doi = {10.48550/arXiv.2503.20858},
archivePrefix = {arXiv},
       eprint = {2503.20858},
 primaryClass = {astro-ph.GA},
       adsurl = {https://ui.adsabs.harvard.edu/abs/2025arXiv250320858F}
}

@ARTICLE{Quenching_Zolotov2015,
       author = {{Zolotov}, Adi and {Dekel}, Avishai and {Mandelker}, Nir and {Tweed}, Dylan and {Inoue}, Shigeki and {DeGraf}, Colin and {Ceverino}, Daniel and {Primack}, Joel R. and {Barro}, Guillermo and {Faber}, Sandra M.},
        title = "{Compaction and quenching of high-z galaxies in cosmological simulations: blue and red nuggets}",
      journal = {\mnras},
         year = 2015,
        month = jul,
       volume = {450},
       number = {3},
        pages = {2327-2353},
          doi = {10.1093/mnras/stv740},
archivePrefix = {arXiv},
       eprint = {1412.4783},
 primaryClass = {astro-ph.GA},
       adsurl = {https://ui.adsabs.harvard.edu/abs/2015MNRAS.450.2327Z}
}

@ARTICLE{SFGGas_Tacchella2016,
       author = {{Tacchella}, Sandro and {Dekel}, Avishai and {Carollo}, C. Marcella and {Ceverino}, Daniel and {DeGraf}, Colin and {Lapiner}, Sharon and {Mandelker}, Nir and {Primack Joel}, R.},
        title = "{The confinement of star-forming galaxies into a main sequence through episodes of gas compaction, depletion and replenishment}",
      journal = {\mnras},
         year = 2016,
        month = apr,
       volume = {457},
       number = {3},
        pages = {2790-2813},
          doi = {10.1093/mnras/stw131},
archivePrefix = {arXiv},
       eprint = {1509.02529},
 primaryClass = {astro-ph.GA},
       adsurl = {https://ui.adsabs.harvard.edu/abs/2016MNRAS.457.2790T}
}

@ARTICLE{EnciWang_and_Lilly2020a,
       author = {{Wang}, Enci and {Lilly}, Simon J.},
        title = "{The Variability of the Star Formation Rate in Galaxies. I. Star Formation Histories Traced by EW(H{\ensuremath{\alpha}}) and EW(H{\ensuremath{\delta}}$_{A}$)}",
      journal = {\apj},
         year = 2020,
        month = apr,
       volume = {892},
       number = {2},
          eid = {87},
        pages = {87},
          doi = {10.3847/1538-4357/ab7b7d},
archivePrefix = {arXiv},
       eprint = {1912.06523},
 primaryClass = {astro-ph.GA},
       adsurl = {https://ui.adsabs.harvard.edu/abs/2020ApJ...892...87W}
}

@ARTICLE{WangSen2025,
       author = {{Wang}, Sen and {Xu}, Dandan and {Lu}, Shengdong},
        title = "{From Larger-scale Cold-gas Angular Momentum Environments to Galaxy Star Formation Activity}",
      journal = {\apj},
         year = 2025,
        month = jun,
       volume = {986},
       number = {1},
          eid = {85},
        pages = {85},
          doi = {10.3847/1538-4357/add155},
archivePrefix = {arXiv},
       eprint = {2411.16849},
 primaryClass = {astro-ph.GA},
       adsurl = {https://ui.adsabs.harvard.edu/abs/2025ApJ...986...85W}
}

@ARTICLE{JennyWan2025,
       author = {{Wan}, Jenny T. and {Tacchella}, Sandro and {D'Eugenio}, Francesco and {Johnson}, Benjamin D. and {van der Wel}, Arjen},
        title = "{Decoding the variability in the star formation histories of z {\ensuremath{\sim}} 0.8 galaxies}",
      journal = {\mnras},
         year = 2025,
        month = jun,
       volume = {539},
       number = {4},
        pages = {2891-2909},
          doi = {10.1093/mnras/staf657},
archivePrefix = {arXiv},
       eprint = {2504.05281},
 primaryClass = {astro-ph.GA},
       adsurl = {https://ui.adsabs.harvard.edu/abs/2025MNRAS.539.2891W}
}

@ARTICLE{MarissaPerry2025,
       author = {{Perry}, Marissa N. and {Taylor}, Anthony J. and {Chavez Ortiz}, Oscar A. and {Finkelstein}, Steven L. and {Leung}, Gene C.~K. and {Bagley}, Micaela B. and {Fernandez}, Vital and {Arrabal Haro}, Pablo and {Chworowsky}, Katherine and {Cleri}, Nikko J. and {Dickinson}, Mark and {Ellis}, Richard S. and {Kartaltepe}, Jeyhan S. and {Koekemoer}, Anton M. and {Pacucci}, Fabio and {Papovich}, Casey and {Pirzkal}, Nor and {Tacchella}, Sandro},
        title = "{The Prevalence of Bursty Star Formation in Low-Mass Galaxies at z=1-7 from H{\ensuremath{\alpha}}-to-UV Diagnostics}",
      journal = {arXiv e-prints},
         year = 2025,
        month = oct,
          eid = {arXiv:2510.05388},
        pages = {arXiv:2510.05388},
          doi = {10.48550/arXiv.2510.05388},
archivePrefix = {arXiv},
       eprint = {2510.05388},
 primaryClass = {astro-ph.GA},
       adsurl = {https://ui.adsabs.harvard.edu/abs/2025arXiv251005388P}
}

@ARTICLE{Tolstoy2009ARAA,
       author = {{Tolstoy}, Eline and {Hill}, Vanessa and {Tosi}, Monica},
        title = "{Star-Formation Histories, Abundances, and Kinematics of Dwarf Galaxies in the Local Group}",
      journal = {\araa},
         year = 2009,
        month = sep,
       volume = {47},
       number = {1},
        pages = {371-425},
          doi = {10.1146/annurev-astro-082708-101650},
archivePrefix = {arXiv},
       eprint = {0904.4505},
 primaryClass = {astro-ph.CO},
       adsurl = {https://ui.adsabs.harvard.edu/abs/2009ARA&A..47..371T}
}

@ARTICLE{McQuinn2010a,
       author = {{McQuinn}, Kristen B.~W. and {Skillman}, Evan D. and {Cannon}, John M. and {Dalcanton}, Julianne and {Dolphin}, Andrew and {Hidalgo-Rodr{\'\i}guez}, Sebastian and {Holtzman}, Jon and {Stark}, David and {Weisz}, Daniel and {Williams}, Benjamin},
        title = "{The Nature of Starbursts. I. The Star Formation Histories of Eighteen Nearby Starburst Dwarf Galaxies}",
      journal = {\apj},
         year = 2010,
        month = sep,
       volume = {721},
       number = {1},
        pages = {297-317},
          doi = {10.1088/0004-637X/721/1/297},
archivePrefix = {arXiv},
       eprint = {1008.1589},
 primaryClass = {astro-ph.CO},
       adsurl = {https://ui.adsabs.harvard.edu/abs/2010ApJ...721..297M}
}

@ARTICLE{McQuinn2010b,
       author = {{McQuinn}, Kristen B.~W. and {Skillman}, Evan D. and {Cannon}, John M. and {Dalcanton}, Julianne and {Dolphin}, Andrew and {Hidalgo-Rodr{\'\i}guez}, Sebastian and {Holtzman}, Jon and {Stark}, David and {Weisz}, Daniel and {Williams}, Benjamin},
        title = "{The Nature of Starbursts. II. The Duration of Starbursts in Dwarf Galaxies}",
      journal = {\apj},
         year = 2010,
        month = nov,
       volume = {724},
       number = {1},
        pages = {49-58},
          doi = {10.1088/0004-637X/724/1/49},
archivePrefix = {arXiv},
       eprint = {1009.2940},
 primaryClass = {astro-ph.CO},
       adsurl = {https://ui.adsabs.harvard.edu/abs/2010ApJ...724...49M}
}

@ARTICLE{Weisz2011,
       author = {{Weisz}, Daniel R. and {Dalcanton}, Julianne J. and {Williams}, Benjamin F. and {Gilbert}, Karoline M. and {Skillman}, Evan D. and {Seth}, Anil C. and {Dolphin}, Andrew E. and {McQuinn}, Kristen B.~W. and {Gogarten}, Stephanie M. and {Holtzman}, Jon and {Rosema}, Keith and {Cole}, Andrew and {Karachentsev}, Igor D. and {Zaritsky}, Dennis},
        title = "{The ACS Nearby Galaxy Survey Treasury. VIII. The Global Star Formation Histories of 60 Dwarf Galaxies in the Local Volume}",
      journal = {\apj},
         year = 2011,
        month = sep,
       volume = {739},
       number = {1},
          eid = {5},
        pages = {5},
          doi = {10.1088/0004-637X/739/1/5},
archivePrefix = {arXiv},
       eprint = {1101.1093},
 primaryClass = {astro-ph.CO},
       adsurl = {https://ui.adsabs.harvard.edu/abs/2011ApJ...739....5W}
}

@ARTICLE{JunYin2023,
       author = {{Yin}, Jun and {Shen}, Shiyin and {Hao}, Lei},
        title = "{Linking the Metallicity Enrichment History to the Star Formation History: An SFH-regulated Chemical Evolution Model and Its Implications for the Gas Cycling Process}",
      journal = {\apj},
         year = 2023,
        month = nov,
       volume = {958},
       number = {1},
          eid = {34},
        pages = {34},
          doi = {10.3847/1538-4357/acfa6b},
archivePrefix = {arXiv},
       eprint = {2310.06785},
 primaryClass = {astro-ph.GA},
       adsurl = {https://ui.adsabs.harvard.edu/abs/2023ApJ...958...34Y}
}

@ARTICLE{Cenci2024,
       author = {{Cenci}, Elia and {Feldmann}, Robert and {Gensior}, Jindra and {Moreno}, Jorge and {Bassini}, Luigi and {Bernardini}, Mauro},
        title = "{Starbursts driven by central gas compaction}",
      journal = {\mnras},
         year = 2024,
        month = jan,
       volume = {527},
       number = {3},
        pages = {7871-7890},
          doi = {10.1093/mnras/stad3709},
archivePrefix = {arXiv},
       eprint = {2309.09046},
 primaryClass = {astro-ph.GA},
       adsurl = {https://ui.adsabs.harvard.edu/abs/2024MNRAS.527.7871C}
}

@ARTICLE{2025MunozLopez,
       author = {{Mu{\~n}oz L{\'o}pez}, C. and {Krajnovi{\'c}}, Davor and {Epinat}, B. and {Urrutia}, T. and {Pessa}, I. and {Contini}, T. and {Nanayakkara}, T. and {Pharo}, J. and {Gon{\c{c}}alves}, T.~S. and {Thai}, Tran Thi and {Bouch{\'e}}, N.~F.},
        title = "{Multiple star-forming episodes of intermediate-redshift galaxies}",
      journal = {arXiv e-prints},
         year = 2025,
        month = sep,
          eid = {arXiv:2509.01710},
        pages = {arXiv:2509.01710},
          doi = {10.48550/arXiv.2509.01710},
archivePrefix = {arXiv},
       eprint = {2509.01710},
 primaryClass = {astro-ph.GA},
       adsurl = {https://ui.adsabs.harvard.edu/abs/2025arXiv250901710M}
}

@ARTICLE{Gui2025,
       author = {{Gui}, Yuqian and {Xu}, Dandan and {Wang}, Haoyi and {Mei}, Xuelun and {Wang}, Enci and {Li}, Cheng and {Wuyts}, Stijn},
        title = "{Episodic Star Formation -- I. Overview and Scatter of the Star-Forming Main Sequence}",
      journal = {arXiv e-prints},
         year = 2025,
        month = nov,
          eid = {arXiv:2512.00151},
        pages = {arXiv:2512.00151},
          doi = {10.48550/arXiv.2512.00151},
archivePrefix = {arXiv},
       eprint = {2512.00151},
 primaryClass = {astro-ph.GA},
       adsurl = {https://ui.adsabs.harvard.edu/abs/2025arXiv251200151G}
}

@ARTICLE{Springel2010,
       author = {{Springel}, Volker},
        title = "{E pur si muove: Galilean-invariant cosmological hydrodynamical simulations on a moving mesh}",
      journal = {\mnras},
         year = 2010,
        month = jan,
       volume = {401},
       number = {2},
        pages = {791-851},
          doi = {10.1111/j.1365-2966.2009.15715.x},
archivePrefix = {arXiv},
       eprint = {0901.4107},
 primaryClass = {astro-ph.CO},
       adsurl = {https://ui.adsabs.harvard.edu/abs/2010MNRAS.401..791S}
}

@ARTICLE{Meert2015,
       author = {{Meert}, Alan and {Vikram}, Vinu and {Bernardi}, Mariangela},
        title = "{A catalogue of 2D photometric decompositions in the SDSS-DR7 spectroscopic main galaxy sample: preferred models and systematics}",
      journal = {\mnras},
         year = 2015,
        month = feb,
       volume = {446},
       number = {4},
        pages = {3943-3974},
          doi = {10.1093/mnras/stu2333},
archivePrefix = {arXiv},
       eprint = {1406.4179},
 primaryClass = {astro-ph.GA},
       adsurl = {https://ui.adsabs.harvard.edu/abs/2015MNRAS.446.3943M}
}

@ARTICLE{Jia2025,
       author = {{Jia}, Cheng and {Wang}, Enci and {Lyu}, Cheqiu and {Ma}, Chengyu and {Song}, Jie and {Chen}, Yangyao and {Wang}, Kai and {Yu}, Haoran and {Chen}, Zeyu and {Wang}, Jinyang and {Wang}, Yifan and {Kong}, Xu},
        title = "{Potential-driven Metal Cycling: JADES Census of Gas-phase Metallicity for Galaxies at 1 < z < 7}",
      journal = {\apjl},
         year = 2025,
        month = jun,
       volume = {986},
       number = {2},
          eid = {L24},
        pages = {L24},
          doi = {10.3847/2041-8213/addfd9},
archivePrefix = {arXiv},
       eprint = {2504.18820},
 primaryClass = {astro-ph.GA},
       adsurl = {https://ui.adsabs.harvard.edu/abs/2025ApJ...986L..24J}
}

@ARTICLE{Tacchella2020,
       author = {{Tacchella}, Sandro and {Forbes}, John C. and {Caplar}, Neven},
        title = "{Stochastic modelling of star-formation histories II: star-formation variability from molecular clouds and gas inflow}",
      journal = {\mnras},
         year = 2020,
        month = sep,
       volume = {497},
       number = {1},
        pages = {698-725},
          doi = {10.1093/mnras/staa1838},
archivePrefix = {arXiv},
       eprint = {2006.09382},
 primaryClass = {astro-ph.GA},
       adsurl = {https://ui.adsabs.harvard.edu/abs/2020MNRAS.497..698T}
}

@ARTICLE{Mannucci2010,
       author = {{Mannucci}, F. and {Cresci}, G. and {Maiolino}, R. and {Marconi}, A. and {Gnerucci}, A.},
        title = "{A fundamental relation between mass, star formation rate and metallicity in local and high-redshift galaxies}",
      journal = {\mnras},
         year = 2010,
        month = nov,
       volume = {408},
       number = {4},
        pages = {2115-2127},
          doi = {10.1111/j.1365-2966.2010.17291.x},
archivePrefix = {arXiv},
       eprint = {1005.0006},
 primaryClass = {astro-ph.CO},
       adsurl = {https://ui.adsabs.harvard.edu/abs/2010MNRAS.408.2115M}
}

@ARTICLE{Shi2011,
       author = {{Shi}, Yong and {Helou}, George and {Yan}, Lin and {Armus}, Lee and {Wu}, Yanling and {Papovich}, Casey and {Stierwalt}, Sabrina},
        title = "{Extended Schmidt Law: Role of Existing Stars in Current Star Formation}",
      journal = {\apj},
         year = 2011,
        month = jun,
       volume = {733},
       number = {2},
          eid = {87},
        pages = {87},
          doi = {10.1088/0004-637X/733/2/87},
archivePrefix = {arXiv},
       eprint = {1103.3711},
 primaryClass = {astro-ph.CO},
       adsurl = {https://ui.adsabs.harvard.edu/abs/2011ApJ...733...87S}
}

@ARTICLE{Wang2018,
       author = {{Wang}, Enci and {Kong}, Xu and {Pan}, Zhizheng},
        title = "{Connecting Compact Star-forming and Extended Star-forming Galaxies at Low Redshift: Implications for Galaxy Compaction and Quenching}",
      journal = {\apj},
         year = 2018,
        month = sep,
       volume = {865},
       number = {1},
          eid = {49},
        pages = {49},
          doi = {10.3847/1538-4357/aadb9e},
archivePrefix = {arXiv},
       eprint = {1808.05929},
 primaryClass = {astro-ph.GA},
       adsurl = {https://ui.adsabs.harvard.edu/abs/2018ApJ...865...49W}
}

@ARTICLE{Ma2024,
       author = {{Ma}, Chengyu and {Wang}, Kai and {Wang}, Enci and {Peng}, Yingjie and {Jiang}, Haochen and {Yu}, Haoran and {Jia}, Cheng and {Chen}, Zeyu and {Li}, Haixin and {Kong}, Xu},
        title = "{Revisiting the Fundamental Metallicity Relation with Observation and Simulation}",
      journal = {\apjl},
         year = 2024,
        month = aug,
       volume = {971},
       number = {1},
          eid = {L14},
        pages = {L14},
          doi = {10.3847/2041-8213/ad675f},
archivePrefix = {arXiv},
       eprint = {2407.21716},
 primaryClass = {astro-ph.GA},
       adsurl = {https://ui.adsabs.harvard.edu/abs/2024ApJ...971L..14M}
}

@ARTICLE{Song2025,
       author = {{Song}, Jie and {Wang}, Enci and {Jia}, Cheng and {Lyu}, Cheqiu and {Chen}, Yangyao and {Wang}, Jinyang and {Li}, Fujia and {Ding}, Weiyu and {Fang}, Guanwen and {Kong}, Xu},
        title = "{Transition from Outside-in to Inside-Out at $z\sim 2$: Evidence from Radial Profiles of Specific Star Formation Rate based on JWST/HST}",
      journal = {arXiv e-prints},
         year = 2025,
        month = dec,
          eid = {arXiv:2512.01684},
        pages = {arXiv:2512.01684},
          doi = {10.48550/arXiv.2512.01684},
archivePrefix = {arXiv},
       eprint = {2512.01684},
 primaryClass = {astro-ph.GA},
       adsurl = {https://ui.adsabs.harvard.edu/abs/2025arXiv251201684S}
}

@ARTICLE{Jia2024,
       author = {{Jia}, Cheng and {Wang}, Enci and {Wang}, Huiyuan and {Li}, Hui and {Yao}, Yao and {Song}, Jie and {Zhang}, Hongxin and {Rong}, Yu and {Chen}, Yangyao and {Yu}, Haoran and {Chen}, Zeyu and {Li}, Haixin and {Ma}, Chengyu and {Kong}, Xu},
        title = "{Size Growth on Short Timescales of Star-forming Galaxies: Insights from Size Variation with Rest-frame Wavelength with JADES}",
      journal = {\apj},
         year = 2024,
        month = dec,
       volume = {977},
       number = {2},
          eid = {165},
        pages = {165},
          doi = {10.3847/1538-4357/ad919a},
archivePrefix = {arXiv},
       eprint = {2411.07458},
 primaryClass = {astro-ph.GA},
       adsurl = {https://ui.adsabs.harvard.edu/abs/2024ApJ...977..165J}
}

@ARTICLE{Wang2022,
       author = {{Wang}, Enci and {Lilly}, Simon J.},
        title = "{The Origin of Exponential Star-forming Disks}",
      journal = {\apj},
         year = 2022,
        month = mar,
       volume = {927},
       number = {2},
          eid = {217},
        pages = {217},
          doi = {10.3847/1538-4357/ac49ed},
archivePrefix = {arXiv},
       eprint = {2201.04148},
 primaryClass = {astro-ph.GA},
       adsurl = {https://ui.adsabs.harvard.edu/abs/2022ApJ...927..217W}
}

@ARTICLE{Wang2021,
       author = {{Wang}, Enci and {Lilly}, Simon J.},
        title = "{Gas-phase Metallicity as a Diagnostic of the Drivers of Star Formation on Different Spatial Scales}",
      journal = {\apj},
         year = 2021,
        month = apr,
       volume = {910},
       number = {2},
          eid = {137},
        pages = {137},
          doi = {10.3847/1538-4357/abe413},
archivePrefix = {arXiv},
       eprint = {2009.01935},
 primaryClass = {astro-ph.GA},
       adsurl = {https://ui.adsabs.harvard.edu/abs/2021ApJ...910..137W}
}

@ARTICLE{Wang2019,
       author = {{Wang}, Enci and {Lilly}, Simon J. and {Pezzulli}, Gabriele and {Matthee}, Jorryt},
        title = "{On the Elevation and Suppression of Star Formation within Galaxies}",
      journal = {\apj},
         year = 2019,
        month = jun,
       volume = {877},
       number = {2},
          eid = {132},
        pages = {132},
          doi = {10.3847/1538-4357/ab1c5b},
archivePrefix = {arXiv},
       eprint = {1901.10276},
 primaryClass = {astro-ph.GA},
       adsurl = {https://ui.adsabs.harvard.edu/abs/2019ApJ...877..132W}
}

@ARTICLE{Peebles_1969,
       author = {{Peebles}, P.~J.~E.},
        title = "{Origin of the Angular Momentum of Galaxies}",
      journal = {\apj},
         year = 1969,
        month = feb,
       volume = {155},
        pages = {393},
          doi = {10.1086/149876},
       adsurl = {https://ui.adsabs.harvard.edu/abs/1969ApJ...155..393P}
}

@ARTICLE{doroshkevich1970,
       author = {{Doroshkevich}, A.~G.},
        title = "{Spatial structure of perturbations and origin of galactic rotation in fluctuation theory}",
      journal = {Astrophysics},
         year = 1970,
        month = oct,
       volume = {6},
       number = {4},
        pages = {320-330},
          doi = {10.1007/BF01001625},
       adsurl = {https://ui.adsabs.harvard.edu/abs/1970Ap......6..320D}
}

@ARTICLE{white1984,
       author = {{White}, S.~D.~M.},
        title = "{Angular momentum growth in protogalaxies}",
      journal = {\apj},
         year = 1984,
        month = nov,
       volume = {286},
        pages = {38-41},
          doi = {10.1086/162573},
       adsurl = {https://ui.adsabs.harvard.edu/abs/1984ApJ...286...38W}
}

@ARTICLE{Fall_Efstathiou_1980,
       author = {{Fall}, S.~M. and {Efstathiou}, G.},
        title = "{Formation and rotation of disc galaxies with haloes.}",
      journal = {\mnras},
         year = 1980,
        month = oct,
       volume = {193},
        pages = {189-206},
          doi = {10.1093/mnras/193.2.189},
       adsurl = {https://ui.adsabs.harvard.edu/abs/1980MNRAS.193..189F}
}

@ARTICLE{Catelan_Theuns_1996,
       author = {{Catelan}, Paolo and {Theuns}, Tom},
        title = "{Non-linear evolution of the angular momentum of protostructures from tidal torques}",
      journal = {\mnras},
         year = 1996,
        month = sep,
       volume = {282},
       number = {2},
        pages = {455-469},
          doi = {10.1093/mnras/282.2.455},
archivePrefix = {arXiv},
       eprint = {astro-ph/9604078},
 primaryClass = {astro-ph},
       adsurl = {https://ui.adsabs.harvard.edu/abs/1996MNRAS.282..455C}
}

@ARTICLE{barnes1987,
       author = {{Barnes}, Joshua and {Efstathiou}, George},
        title = "{Angular Momentum from Tidal Torques}",
      journal = {\apj},
         year = 1987,
        month = aug,
       volume = {319},
        pages = {575},
          doi = {10.1086/165480},
       adsurl = {https://ui.adsabs.harvard.edu/abs/1987ApJ...319..575B}
}

@ARTICLE{Schaefer_2009_Review,
       author = {{Sch{\"a}fer}, Bj{\"o}rn Malte},
        title = "{Galactic Angular Momenta and Angular Momentum Correlations in the Cosmological Large-Scale Structure}",
      journal = {International Journal of Modern Physics D},
         year = 2009,
        month = jan,
       volume = {18},
       number = {2},
        pages = {173-222},
          doi = {10.1142/S0218271809014388},
archivePrefix = {arXiv},
       eprint = {0808.0203},
 primaryClass = {astro-ph},
       adsurl = {https://ui.adsabs.harvard.edu/abs/2009IJMPD..18..173S}
}

@ARTICLE{Wuyts2011,
       author = {{Wuyts}, Stijn and {F{\"o}rster Schreiber}, Natascha M. and {van der Wel}, Arjen and {Magnelli}, Benjamin and {Guo}, Yicheng and {Genzel}, Reinhard and {Lutz}, Dieter and {Aussel}, Herv{\'e} and {Barro}, Guillermo and {Berta}, Stefano and {Cava}, Antonio and {Graci{\'a}-Carpio}, Javier and {Hathi}, Nimish P. and {Huang}, Kuang-Han and {Kocevski}, Dale D. and {Koekemoer}, Anton M. and {Lee}, Kyoung-Soo and {Le Floc'h}, Emeric and {McGrath}, Elizabeth J. and {Nordon}, Raanan and {Popesso}, Paola and {Pozzi}, Francesca and {Riguccini}, Laurie and {Rodighiero}, Giulia and {Saintonge}, Amelie and {Tacconi}, Linda},
        title = "{Galaxy Structure and Mode of Star Formation in the SFR-Mass Plane from z \raisebox{-0.5ex}\textasciitilde 2.5 to z \raisebox{-0.5ex}\textasciitilde 0.1}",
      journal = {\apj},
         year = 2011,
        month = dec,
       volume = {742},
       number = {2},
          eid = {96},
        pages = {96},
          doi = {10.1088/0004-637X/742/2/96},
archivePrefix = {arXiv},
       eprint = {1107.0317},
 primaryClass = {astro-ph.CO},
       adsurl = {https://ui.adsabs.harvard.edu/abs/2011ApJ...742...96W}
}

@ARTICLE{Barro2019,
       author = {{Barro}, Guillermo and {P{\'e}rez-Gonz{\'a}lez}, Pablo G. and {Cava}, Antonio and {Brammer}, Gabriel and {Pandya}, Viraj and {Eliche Moral}, Carmen and {Esquej}, Pilar and {Dom{\'\i}nguez-S{\'a}nchez}, Helena and {Alcalde Pampliega}, Belen and {Guo}, Yicheng and {Koekemoer}, Anton M. and {Trump}, Jonathan R. and {Ashby}, Matthew L.~N. and {Cardiel}, Nicolas and {Castellano}, Marco and {Conselice}, Christopher J. and {Dickinson}, Mark E. and {Dolch}, Timothy and {Donley}, Jennifer L. and {Espino Briones}, N{\'e}stor and {Faber}, Sandra M. and {Fazio}, Giovanni G. and {Ferguson}, Henry and {Finkelstein}, Steve and {Fontana}, Adriano and {Galametz}, Audrey and {Gardner}, Jonathan P. and {Gawiser}, Eric and {Giavalisco}, Mauro and {Grazian}, Andrea and {Grogin}, Norman A. and {Hathi}, Nimish P. and {Hemmati}, Shoubaneh and {Hern{\'a}n-Caballero}, Antonio and {Kocevski}, Dale and {Koo}, David C. and {Kodra}, Dritan and {Lee}, Kyoung-Soo and {Lin}, Lihwai and {Lucas}, Ray A. and {Mobasher}, Bahram and {McGrath}, Elizabeth J. and {Nandra}, Kirpal and {Nayyeri}, Hooshang and {Newman}, Jeffrey A. and {Pforr}, Janine and {Peth}, Michael and {Rafelski}, Marc and {Rodr{\'\i}guez-Munoz}, Lucia and {Salvato}, Mara and {Stefanon}, Mauro and {van der Wel}, Arjen and {Willner}, Steven P. and {Wiklind}, Tommy and {Wuyts}, Stijn},
        title = "{The CANDELS/SHARDS Multiwavelength Catalog in GOODS-N: Photometry, Photometric Redshifts, Stellar Masses, Emission-line Fluxes, and Star Formation Rates}",
      journal = {\apjs},
         year = 2019,
        month = aug,
       volume = {243},
       number = {2},
          eid = {22},
        pages = {22},
          doi = {10.3847/1538-4365/ab23f2},
archivePrefix = {arXiv},
       eprint = {1908.00569},
 primaryClass = {astro-ph.GA},
       adsurl = {https://ui.adsabs.harvard.edu/abs/2019ApJS..243...22B}
}

@ARTICLE{Kirek2009FAST,
       author = {{Kriek}, Mariska and {van Dokkum}, Pieter G. and {Franx}, Marijn and {Illingworth}, Garth D. and {Magee}, Daniel K.},
        title = "{The Hubble Sequence Beyond z = 2 for Massive Galaxies: Contrasting Large Star-forming and Compact Quiescent Galaxies}",
      journal = {\apjl},
         year = 2009,
        month = nov,
       volume = {705},
       number = {1},
        pages = {L71-L75},
          doi = {10.1088/0004-637X/705/1/L71},
archivePrefix = {arXiv},
       eprint = {0909.0260},
 primaryClass = {astro-ph.CO},
       adsurl = {https://ui.adsabs.harvard.edu/abs/2009ApJ...705L..71K}
}

@ARTICLE{SDSSDR7,
       author = {{Abazajian}, Kevork N. and {Adelman-McCarthy}, Jennifer K. and {Ag{\"u}eros}, Marcel A. and {Allam}, Sahar S. and {Allende Prieto}, Carlos and {An}, Deokkeun and {Anderson}, Kurt S.~J. and {Anderson}, Scott F. and {Annis}, James and {Bahcall}, Neta A. and {Bailer-Jones}, C.~A.~L. and {Barentine}, J.~C. and {Bassett}, Bruce A. and {Becker}, Andrew C. and {Beers}, Timothy C. and {Bell}, Eric F. and {Belokurov}, Vasily and {Berlind}, Andreas A. and {Berman}, Eileen F. and {Bernardi}, Mariangela and {Bickerton}, Steven J. and {Bizyaev}, Dmitry and {Blakeslee}, John P. and {Blanton}, Michael R. and {Bochanski}, John J. and {Boroski}, William N. and {Brewington}, Howard J. and {Brinchmann}, Jarle and {Brinkmann}, J. and {Brunner}, Robert J. and {Budav{\'a}ri}, Tam{\'a}s and {Carey}, Larry N. and {Carliles}, Samuel and {Carr}, Michael A. and {Castander}, Francisco J. and {Cinabro}, David and {Connolly}, A.~J. and {Csabai}, Istv{\'a}n and {Cunha}, Carlos E. and {Czarapata}, Paul C. and {Davenport}, James R.~A. and {de Haas}, Ernst and {Dilday}, Ben and {Doi}, Mamoru and {Eisenstein}, Daniel J. and {Evans}, Michael L. and {Evans}, N.~W. and {Fan}, Xiaohui and {Friedman}, Scott D. and {Frieman}, Joshua A. and {Fukugita}, Masataka and {G{\"a}nsicke}, Boris T. and {Gates}, Evalyn and {Gillespie}, Bruce and {Gilmore}, G. and {Gonzalez}, Belinda and {Gonzalez}, Carlos F. and {Grebel}, Eva K. and {Gunn}, James E. and {Gy{\"o}ry}, Zsuzsanna and {Hall}, Patrick B. and {Harding}, Paul and {Harris}, Frederick H. and {Harvanek}, Michael and {Hawley}, Suzanne L. and {Hayes}, Jeffrey J.~E. and {Heckman}, Timothy M. and {Hendry}, John S. and {Hennessy}, Gregory S. and {Hindsley}, Robert B. and {Hoblitt}, J. and {Hogan}, Craig J. and {Hogg}, David W. and {Holtzman}, Jon A. and {Hyde}, Joseph B. and {Ichikawa}, Shin-ichi and {Ichikawa}, Takashi and {Im}, Myungshin and {Ivezi{\'c}}, {\v{Z}}eljko and {Jester}, Sebastian and {Jiang}, Linhua and {Johnson}, Jennifer A. and {Jorgensen}, Anders M. and {Juri{\'c}}, Mario and {Kent}, Stephen M. and {Kessler}, R. and {Kleinman}, S.~J. and {Knapp}, G.~R. and {Konishi}, Kohki and {Kron}, Richard G. and {Krzesinski}, Jurek and {Kuropatkin}, Nikolay and {Lampeitl}, Hubert and {Lebedeva}, Svetlana and {Lee}, Myung Gyoon and {Lee}, Young Sun and {French Leger}, R. and {L{\'e}pine}, S{\'e}bastien and {Li}, Nolan and {Lima}, Marcos and {Lin}, Huan and {Long}, Daniel C. and {Loomis}, Craig P. and {Loveday}, Jon and {Lupton}, Robert H. and {Magnier}, Eugene and {Malanushenko}, Olena and {Malanushenko}, Viktor and {Mandelbaum}, Rachel and {Margon}, Bruce and {Marriner}, John P. and {Mart{\'\i}nez-Delgado}, David and {Matsubara}, Takahiko and {McGehee}, Peregrine M. and {McKay}, Timothy A. and {Meiksin}, Avery and {Morrison}, Heather L. and {Mullally}, Fergal and {Munn}, Jeffrey A. and {Murphy}, Tara and {Nash}, Thomas and {Nebot}, Ada and {Neilsen}, Jr., Eric H. and {Newberg}, Heidi Jo and {Newman}, Peter R. and {Nichol}, Robert C. and {Nicinski}, Tom and {Nieto-Santisteban}, Maria and {Nitta}, Atsuko and {Okamura}, Sadanori and {Oravetz}, Daniel J. and {Ostriker}, Jeremiah P. and {Owen}, Russell and {Padmanabhan}, Nikhil and {Pan}, Kaike and {Park}, Changbom and {Pauls}, George and {Peoples}, Jr., John and {Percival}, Will J. and {Pier}, Jeffrey R. and {Pope}, Adrian C. and {Pourbaix}, Dimitri and {Price}, Paul A. and {Purger}, Norbert and {Quinn}, Thomas and {Raddick}, M. Jordan and {Re Fiorentin}, Paola and {Richards}, Gordon T. and {Richmond}, Michael W. and {Riess}, Adam G. and {Rix}, Hans-Walter and {Rockosi}, Constance M. and {Sako}, Masao and {Schlegel}, David J. and {Schneider}, Donald P. and {Scholz}, Ralf-Dieter and {Schreiber}, Matthias R. and {Schwope}, Axel D. and {Seljak}, Uro{\v{s}} and {Sesar}, Branimir and {Sheldon}, Erin and {Shimasaku}, Kazu and {Sibley}, Valena C. and {Simmons}, A.~E. and {Sivarani}, Thirupathi and {Allyn Smith}, J. and {Smith}, Martin C. and {Smol{\v{c}}i{\'c}}, Vernesa and {Snedden}, Stephanie A. and {Stebbins}, Albert and {Steinmetz}, Matthias and {Stoughton}, Chris and {Strauss}, Michael A. and {SubbaRao}, Mark and {Suto}, Yasushi and {Szalay}, Alexander S. and {Szapudi}, Istv{\'a}n and {Szkody}, Paula and {Tanaka}, Masayuki and {Tegmark}, Max and {Teodoro}, Luis F.~A. and {Thakar}, Aniruddha R. and {Tremonti}, Christy A. and {Tucker}, Douglas L. and {Uomoto}, Alan and {Vanden Berk}, Daniel E. and {Vandenberg}, Jan and {Vidrih}, S. and {Vogeley}, Michael S. and {Voges}, Wolfgang and {Vogt}, Nicole P. and {Wadadekar}, Yogesh and {Watters}, Shannon and {Weinberg}, David H. and {West}, Andrew A. and {White}, Simon D.~M. and {Wilhite}, Brian C. and {Wonders}, Alainna C. and {Yanny}, Brian and {Yocum}, D.~R.},
        title = "{The Seventh Data Release of the Sloan Digital Sky Survey}",
      journal = {\apjs},
         year = 2009,
        month = jun,
       volume = {182},
       number = {2},
        pages = {543-558},
          doi = {10.1088/0067-0049/182/2/543},
archivePrefix = {arXiv},
       eprint = {0812.0649},
 primaryClass = {astro-ph},
       adsurl = {https://ui.adsabs.harvard.edu/abs/2009ApJS..182..543A}
}

@ARTICLE{YangXiaohu2007,
       author = {{Yang}, Xiaohu and {Mo}, H.~J. and {van den Bosch}, Frank C. and {Pasquali}, Anna and {Li}, Cheng and {Barden}, Marco},
        title = "{Galaxy Groups in the SDSS DR4. I. The Catalog and Basic Properties}",
      journal = {\apj},
         year = 2007,
        month = dec,
       volume = {671},
       number = {1},
        pages = {153-170},
          doi = {10.1086/522027},
archivePrefix = {arXiv},
       eprint = {0707.4640},
 primaryClass = {astro-ph},
       adsurl = {https://ui.adsabs.harvard.edu/abs/2007ApJ...671..153Y}
}

@ARTICLE{Salim2017,
       author = {{Salim}, Samir and {Rich}, R. Michael and {Charlot}, St{\'e}phane and {Brinchmann}, Jarle and {Johnson}, Benjamin D. and {Schiminovich}, David and {Seibert}, Mark and {Mallery}, Ryan and {Heckman}, Timothy M. and {Forster}, Karl and {Friedman}, Peter G. and {Martin}, D. Christopher and {Morrissey}, Patrick and {Neff}, Susan G. and {Small}, Todd and {Wyder}, Ted K. and {Bianchi}, Luciana and {Donas}, Jos{\'e} and {Lee}, Young-Wook and {Madore}, Barry F. and {Milliard}, Bruno and {Szalay}, Alex S. and {Welsh}, Barry Y. and {Yi}, Sukyoung K.},
        title = "{UV Star Formation Rates in the Local Universe}",
      journal = {\apjs},
         year = 2007,
        month = dec,
       volume = {173},
       number = {2},
        pages = {267-292},
          doi = {10.1086/519218},
archivePrefix = {arXiv},
       eprint = {0704.3611},
 primaryClass = {astro-ph},
       adsurl = {https://ui.adsabs.harvard.edu/abs/2007ApJS..173..267S}
}

@ARTICLE{Brinchmann2004,
       author = {{Brinchmann}, J. and {Charlot}, S. and {White}, S.~D.~M. and {Tremonti}, C. and {Kauffmann}, G. and {Heckman}, T. and {Brinkmann}, J.},
        title = "{The physical properties of star-forming galaxies in the low-redshift Universe}",
      journal = {\mnras},
         year = 2004,
        month = jul,
       volume = {351},
       number = {4},
        pages = {1151-1179},
          doi = {10.1111/j.1365-2966.2004.07881.x},
archivePrefix = {arXiv},
       eprint = {astro-ph/0311060},
 primaryClass = {astro-ph},
       adsurl = {https://ui.adsabs.harvard.edu/abs/2004MNRAS.351.1151B}
}

@ARTICLE{Donnari2018,
       author = {{Donnari}, Martina and {Pillepich}, Annalisa and {Nelson}, Dylan and {Vogelsberger}, Mark and {Genel}, Shy and {Weinberger}, Rainer and {Marinacci}, Federico and {Springel}, Volker and {Hernquist}, Lars},
        title = "{The star formation activity of IllustrisTNG galaxies: main sequence, UVJ diagram, quenched fractions, and systematics}",
      journal = {\mnras},
         year = 2019,
        month = jun,
       volume = {485},
       number = {4},
        pages = {4817-4840},
          doi = {10.1093/mnras/stz712},
archivePrefix = {arXiv},
       eprint = {1812.07584},
 primaryClass = {astro-ph.GA},
       adsurl = {https://ui.adsabs.harvard.edu/abs/2019MNRAS.485.4817D}
}

@ARTICLE{2019Huertas-Company,
       author = {{Huertas-Company}, Marc and {Rodriguez-Gomez}, Vicente and {Nelson}, Dylan and {Pillepich}, Annalisa and {Bottrell}, Connor and {Bernardi}, Mariangela and {Dom{\'\i}nguez-S{\'a}nchez}, Helena and {Genel}, Shy and {Pakmor}, Ruediger and {Snyder}, Gregory F. and {Vogelsberger}, Mark},
        title = "{The Hubble Sequence at z {\ensuremath{\sim}} 0 in the IllustrisTNG simulation with deep learning}",
      journal = {\mnras},
         year = 2019,
        month = oct,
       volume = {489},
       number = {2},
        pages = {1859-1879},
          doi = {10.1093/mnras/stz2191},
archivePrefix = {arXiv},
       eprint = {1903.07625},
 primaryClass = {astro-ph.GA},
       adsurl = {https://ui.adsabs.harvard.edu/abs/2019MNRAS.489.1859H}
}

@ARTICLE{EAGLE-DR,
       author = {{McAlpine}, S. and {Helly}, J.~C. and {Schaller}, M. and {Trayford}, J.~W. and {Qu}, Y. and {Furlong}, M. and {Bower}, R.~G. and {Crain}, R.~A. and {Schaye}, J. and {Theuns}, T. and {Dalla Vecchia}, C. and {Frenk}, C.~S. and {McCarthy}, I.~G. and {Jenkins}, A. and {Rosas-Guevara}, Y. and {White}, S.~D.~M. and {Baes}, M. and {Camps}, P. and {Lemson}, G.},
        title = "{The EAGLE simulations of galaxy formation: Public release of halo and galaxy catalogues}",
      journal = {Astronomy and Computing},
         year = 2016,
        month = apr,
       volume = {15},
        pages = {72-89},
          doi = {10.1016/j.ascom.2016.02.004},
archivePrefix = {arXiv},
       eprint = {1510.01320},
 primaryClass = {astro-ph.GA},
       adsurl = {https://ui.adsabs.harvard.edu/abs/2016A&C....15...72M}
}

@ARTICLE{Furong2015,
       author = {{Furlong}, M. and {Bower}, R.~G. and {Theuns}, T. and {Schaye}, J. and {Crain}, R.~A. and {Schaller}, M. and {Dalla Vecchia}, C. and {Frenk}, C.~S. and {McCarthy}, I.~G. and {Helly}, J. and {Jenkins}, A. and {Rosas-Guevara}, Y.~M.},
        title = "{Evolution of galaxy stellar masses and star formation rates in the EAGLE simulations}",
      journal = {\mnras},
         year = 2015,
        month = jul,
       volume = {450},
       number = {4},
        pages = {4486-4504},
          doi = {10.1093/mnras/stv852},
archivePrefix = {arXiv},
       eprint = {1410.3485},
 primaryClass = {astro-ph.GA},
       adsurl = {https://ui.adsabs.harvard.edu/abs/2015MNRAS.450.4486F}
}

@ARTICLE{Furong2017,
       author = {{Furlong}, M. and {Bower}, R.~G. and {Crain}, R.~A. and {Schaye}, J. and {Theuns}, T. and {Trayford}, J.~W. and {Qu}, Y. and {Schaller}, M. and {Berthet}, M. and {Helly}, J.~C.},
        title = "{Size evolution of normal and compact galaxies in the EAGLE simulation}",
      journal = {\mnras},
         year = 2017,
        month = feb,
       volume = {465},
       number = {1},
        pages = {722-738},
          doi = {10.1093/mnras/stw2740},
archivePrefix = {arXiv},
       eprint = {1510.05645},
 primaryClass = {astro-ph.GA},
       adsurl = {https://ui.adsabs.harvard.edu/abs/2017MNRAS.465..722F}
}

@ARTICLE{Noeske2007,
       author = {{Noeske}, K.~G. and {Weiner}, B.~J. and {Faber}, S.~M. and {Papovich}, C. and {Koo}, D.~C. and {Somerville}, R.~S. and {Bundy}, K. and {Conselice}, C.~J. and {Newman}, J.~A. and {Schiminovich}, D. and {Le Floc'h}, E. and {Coil}, A.~L. and {Rieke}, G.~H. and {Lotz}, J.~M. and {Primack}, J.~R. and {Barmby}, P. and {Cooper}, M.~C. and {Davis}, M. and {Ellis}, R.~S. and {Fazio}, G.~G. and {Guhathakurta}, P. and {Huang}, J. and {Kassin}, S.~A. and {Martin}, D.~C. and {Phillips}, A.~C. and {Rich}, R.~M. and {Small}, T.~A. and {Willmer}, C.~N.~A. and {Wilson}, G.},
        title = "{Star Formation in AEGIS Field Galaxies since z=1.1: The Dominance of Gradually Declining Star Formation, and the Main Sequence of Star-forming Galaxies}",
      journal = {\apjl},
         year = 2007,
        month = may,
       volume = {660},
       number = {1},
        pages = {L43-L46},
          doi = {10.1086/517926},
archivePrefix = {arXiv},
       eprint = {astro-ph/0701924},
 primaryClass = {astro-ph},
       adsurl = {https://ui.adsabs.harvard.edu/abs/2007ApJ...660L..43N}
}

@ARTICLE{Daddi2007,
       author = {{Daddi}, E. and {Dickinson}, M. and {Morrison}, G. and {Chary}, R. and {Cimatti}, A. and {Elbaz}, D. and {Frayer}, D. and {Renzini}, A. and {Pope}, A. and {Alexander}, D.~M. and {Bauer}, F.~E. and {Giavalisco}, M. and {Huynh}, M. and {Kurk}, J. and {Mignoli}, M.},
        title = "{Multiwavelength Study of Massive Galaxies at z\raisebox{-0.5ex}\textasciitilde2. I. Star Formation and Galaxy Growth}",
      journal = {\apj},
         year = 2007,
        month = nov,
       volume = {670},
       number = {1},
        pages = {156-172},
          doi = {10.1086/521818},
archivePrefix = {arXiv},
       eprint = {0705.2831},
 primaryClass = {astro-ph},
       adsurl = {https://ui.adsabs.harvard.edu/abs/2007ApJ...670..156D}
}

@ARTICLE{Elbaz2007,
       author = {{Elbaz}, D. and {Daddi}, E. and {Le Borgne}, D. and {Dickinson}, M. and {Alexander}, D.~M. and {Chary}, R. -R. and {Starck}, J. -L. and {Brandt}, W.~N. and {Kitzbichler}, M. and {MacDonald}, E. and {Nonino}, M. and {Popesso}, P. and {Stern}, D. and {Vanzella}, E.},
        title = "{The reversal of the star formation-density relation in the distant universe}",
      journal = {\aap},
         year = 2007,
        month = jun,
       volume = {468},
       number = {1},
        pages = {33-48},
          doi = {10.1051/0004-6361:20077525},
archivePrefix = {arXiv},
       eprint = {astro-ph/0703653},
 primaryClass = {astro-ph},
       adsurl = {https://ui.adsabs.harvard.edu/abs/2007A&A...468...33E}
}
\bibliographystyle{aasjournal}



\end{document}